\documentclass[12pt]{article}
\usepackage{soul}

\usepackage[utf8]{inputenc}
\usepackage{graphicx}	% Including figure files
\usepackage{amsmath}	% Advanced maths commands
\usepackage{amssymb}	% Extra maths symbols
\usepackage{xcolor}
\usepackage{microtype}
\usepackage[a4paper, margin=2.4cm]{geometry}
\usepackage{url}
\usepackage{amsmath}
\usepackage{chemformula}
\usepackage{mathptmx}

\usepackage{helvet}
\usepackage[left, pagewise, edtable]{lineno}
\usepackage[font=sf]{caption}

\usepackage{CJK}
\usepackage[whole]{bxcjkjatype}
\usepackage{CJKutf8}
\usepackage{chemgreek}

\def\Msun{M$_\odot$}

\def\kms{km\,s$^{-1}$}

\def\apj{Astrophys.~J.}
\def\apjl{Astrophys.~J.}
\def\apjs{Astrophys.~J.~Suppl.}
\def\mnras{Mon.~Not.~R.~Astron.~Soc.}

\def\nat{Nature}

\def\pasp{Publ.~Astron.~Soc.~Pac.}
\def\aj{Astron.~J.}
\def\aap{Astron.~Astrophys.}
\def\araa{Annu.~Rev.~Astron.~Astrophys.}

\def\apss{Astrophys.~Space~Sci.}
\def\aapr{Astron. Astrophys. Rev.}
\def\aaps{Astron. Astrophys. Suppl. Ser.}
\DeclareRobustCommand{\ion}[2]{%
\relax\ifmmode
\ifx\testbx\f@series
{\mathbf{#1\,\mathsc{#2}}}\else
{\mathrm{#1\,\mathsc{#2}}}\fi
\else\textup{#1\,{\mdseries\textsc{#2}}}%
\fi}

\makeatletter
\renewcommand{\maketitle}{\bgroup\setlength{\parindent}{0pt}
\begin{flushleft}
  \textbf{\LARGE \@title}
  \\
  \vspace{0.5cm}
  \@author
\end{flushleft}\egroup
}
\makeatother

\title{Newly Formed Dust within the Circumstellar Environment of SNIa-CSM 2018evt}
\date{}

\author{\large{Lingzhi Wang\begin{CJK}{UTF8}{gbsn}
(王灵芝)
\end{CJK}$^{1,2,\dagger,*}$,  
Maokai Hu$^{3,\dagger}$,
Lifan Wang$^{4,\dagger}$, 
Yi Yang\begin{CJK}{UTF8}{gbsn}
(杨轶)
\end{CJK}$^{5,6,\ddagger}$, 
Jiawen Yang$^4$,
Haley Gomez$^7$, 
Sijie Chen$^4$,
Lei Hu$^{3,8}$, 
Ting-Wan Chen$^{9}$,
Jun Mo$^5$, 
Xiaofeng Wang$^{5,10}$, 
Dietrich Baade$^{11}$,
Peter Hoeflich$^{12}$, 
J. Craig Wheeler$^{13}$,
Giuliano Pignata$^{14,15}$,
Jamison Burke$^{16,17}$,
Daichi Hiramatsu$^{18,19}$,
D. Andrew Howell$^{16,17}$, 
Curtis McCully$^{16}$,
Craig Pellegrino$^{16,17}$,
Llu{\'i}s Galbany$^{20,21}$,
Eric Y. Hsiao$^{12}$, 
David J. Sand$^{22}$, 
Jujia Zhang$^{23}$,
Syed A. Uddin$^{4}$,
J.~P. Anderson$^{24,15}$,
Chris Ashall$^{25}$, 
Cheng Cheng$^1$,
Mariusz Gromadzki$^{26}$,
Cosimo Inserra $^{7}$,
Han Lin$^5$,
N. Morrell$^{27}$,
Antonia Morales-Garoffolo$^{28}$, 
T. E. M\"uller-Bravo$^{20,21}$,
Matt Nicholl $^{29}$,
Estefania Padilla Gonzalez$^{16,17}$,
M. M. Phillips$^{27}$, 
J. Pineda-Garc\'ia$^{30,15}$, 
Hanna Sai$^5$, 
Mathew Smith $^{31}$,
M. Shahbandeh$^{12}$,
Shubham Srivastav$^{29}$,
M. D. Stritzinger$^{32}$,
Sheng Yang $^{33}$,
D. R. Young $^{29}$,
Lixin Yu$^1$,
Xinghan Zhang$^5$
}
\\
  \vspace{0.5cm}
\normalsize
$^{1}$Chinese Academy of Sciences South America Center for Astronomy (CASSACA), National Astronomical Observatories, CAS, Beijing 100101, China\\
$^{2}$CAS Key Laboratory of Optical Astronomy, National Astronomical Observatories, Chinese Academy of Sciences, Beijing 100101, China\\
$^3$Purple Mountain Observatory, Chinese Academy of Sciences, Nanjing 210023, China\\
$^4$George P. and Cynthia Woods Mitchell Institute for Fundamental Physics and Astronomy, Texas A\&M University, Department of Physics and Astronomy, College Station, TX 77843, USA\\
$^5$Physics Department and Tsinghua Center for Astrophysics (THCA), Tsinghua University, Beijing, $100084$, China\\
$^6$Department of Astronomy, University of California, Berkeley, CA 94720-3411, USA \\
$^{7}$Cardiff Hub for Astrophysics Research and Technology, School of Physics \& Astronomy, Cardiff University, Queens Buildings, The Parade, Cardiff, CF24 3AA, UK\\
$^8$McWilliams Center for Cosmology, Department of Physics, Carnegie Mellon University, Pittsburgh, PA, USA\\
$^{9}$Graduate Institute of Astronomy, National Central University, 300 Jhongda Road, 32001 Jhongli, Taiwan\\
$^{10}$Beijing Planetarium, Beijing Academy of Science and Technology, Beijing, 100044, China\\
$^{11}$European Organisation for Astronomical Research in the Southern Hemisphere (ESO), Karl-Schwarzschild-Str. 2, 85748 Garching b. München, Germany \\
$^{12}$Department of Physics, Florida State University, Tallahassee, Florida 32306-4350, USA\\
$^{13}$Department of Astronomy, University of Texas, Austin, TX 78712, USA \\
$^{14}$Instituto de Alta Investigaci\'on, Universidad de Tarapac\'a, Casilla 7D, Arica, Chile\\
$^{15}$Millennium Institute of Astrophysics (MAS), Nuncio Monse\~nor S\'otero Sanz 100, Providencia, Santiago, Chile\\
$^{16}$Las Cumbres Observatory, 6740 Cortona Drive, Suite 102, Goleta, CA 93117-5575, USA\\
$^{17}$Department of Physics, University of California, Santa Barbara, CA 93106-9530, USA\\
$^{18}$Center for Astrophysics \textbar{} Harvard \& Smithsonian, 60 Garden Street, Cambridge, MA 02138-1516, USA\\
$^{19}$The NSF AI Institute for Artificial Intelligence and Fundamental Interactions\\
$^{20}$Institute of Space Sciences (ICE, CSIC), Campus UAB, Carrer de Can Magrans, s/n, E-08193 Barcelona, Spain\\
$^{21}$Institut d'Estudis Espacials de Catalunya (IEEC), E-08034 Barcelona, Spain\\
$^{22}$Department of Astronomy and Steward Observatory, University of Arizona, 933 North Cherry Avenue, Tucson, AZ 85721-0065, USA\\
$^{23}$Yunnan Observatories, Chinese Academy of Sciences, Kunming 650216, China\\
$^{24}$European Southern Observatory, Alonso de C\'ordova 3107, Casilla 19, Santiago, Chile\\
$^{25}$Department of Physics, Virginia Tech, Blacksburg, VA 24061, USA \\
$^{26}$Astronomical Observatory, University of Warsaw, Al. Ujazdowskie 4, 00-478 Warszawa, Poland \\
$^{27}$Carnegie Observatories, Las Campanas Observatory, Casilla 601, La Serena, Chile \\
$^{28}$Department of Applied Physics, School of Engineering, University of C\'adiz, Campus of Puerto Real, E-11519 C\'adiz, Spain\\
$^{29}$Astrophysics Research Centre, School of Mathematics and Physics, Queen's University Belfast, Belfast BT7 1NN, UK\\
$^{30}$Departamento de Ciencias F\'isicas, Universidad Andres Bello, Avda. Rep\'ublica 252, Santiago, 8320000, Chile\\
$^{31}$Universit\'e de Lyon, Universit\'e Claude Bernard Lyon 1, CNRS/IN2P3, IP2I Lyon, UMR 5822, F-69622, Villeurbanne, France\\
$^{32}$Department of Physics and Astronomy, Aarhus University, Ny Munkegade 120, DK-8000 Aarhus C, Denmark\\
$^{33}$Henan Academy of Sciences, Zhengzhou 450046, Henan, China\\
$^\dagger$ These authors contributed equally\\
$^\ddagger${Bengier-Winslow-Robertson Postdoctoral Fellow} \\
$^*$ Correspondence author: Lingzhi Wang ({\it wanglingzhi@bao.ac.cn}).
}
\begin{document}
%\linenumbers
\maketitle
\noindent\textbf{
Dust associated with various stellar sources in galaxies at all cosmic epochs remains a controversial topic, particularly whether supernovae (SNe) play an important role in dust production.
We report evidence of dust formation in the cold, dense shell behind the ejecta-circumstellar medium (CSM) interaction in the Type Ia-CSM SN 2018evt three years after the explosion, characterized by a rise in the mid-infrared (MIR) emission accompanied by an accelerated decline in the optical radiation of the SN. Such a dust-formation picture is also corroborated by the concurrent evolution of the profiles of the H$\alpha$ emission line. 
Our model suggests enhanced CSM dust concentration at increasing distances from the SN as compared to what can be expected from the density profile of the mass loss from a steady stellar wind. 
By the time of the last MIR observations at day $+$1041, a total amount of $1.2\pm0.2 \times 10^{-2}$\, M$_{\odot}$ of new dust has been formed by SN\,2018evt, making SN~2018evt one of the most prolific dust factories among SNe with evidence of dust formation. The unprecedented witness of the intense production procedure of dust may shed light on the perceptions of dust formation in cosmic history. 
}

\medskip
\section{Main}
\subsection{Introduction}

The content and species of dust grains that are associated with stellar sources in galaxies at all cosmic epochs remain a controversial topic, particularly whether supernovae play an important role in dust production. 
Moreover, they may even carve dust-hostile environments \cite{Morgan_Edmunds03, Matsuura09}, considering ambient grains in any outflow of stellar wind of the SN progenitor may become immediately sublimated and destroyed by the energetic radiation pulse produced by the SN explosion \cite{Hoflich95:DDT, Ferrara_Peroux21}. 
To date, freshly-formed dust has been observed in a handful of core-collapse (CC) SNe, both in the ejecta in-situ \cite{Rho08, Barlow10, Matsuura11, Gomez12a} and its interaction zone with CSM (e.g., \cite{Rho08, Smith09:05ip, Gall14}). 
No clear observational evidence thus far shows any major formation process of dust grains in the thermonuclear runaway of $\sim$1\,M$_{\odot}$ carbon/oxygen white dwarfs (C/O WDs; \cite{Gerardy07:Ia_MIR,Gomez12b}). 

SNe Ia are generally thought to result from thermonuclear explosions of a white dwarf in a binary system. A rare subclass of SNe\,Ia is denoted as SN Ia-CSM, which is thought to be an exploding WD surrounded by a substantial amount of CSM \cite{Silverman13:Iacsm}.
The spectra of such events near peak luminosity are characterized by narrow Balmer emission lines superimposed together with relatively shallow Fe-group and intermediate-mass elements. 
SN~2002ic was the first reported case of SNIa-CSM that revealed large amounts of CSM seen as a strong hydrogen emission \cite{Hamuy03:sn2002ic,Wang04,Deng04,WoodVasey04}.
A number of additional SNIa-CSM have been discovered and studied in detail in the past years, which include SNe 2005gj, PTF11kx, 2012ca, 2013dn, 2015cp, and a recent sample of ZTF SNe ( see \cite{Sharma23} and references therein). 

SN 2018evt (ASASSN-18ro \cite{discover:18evt}) is an {\bf SNIa-CSM} found in the spiral galaxy MCG-01-35-011 at redshift z = 0.02523 \cite{Yang22}.
SN~2018evt shares some common optical spectral features as typical Type Ia SN~1991T-like SNe as shown in {\bf Extended Data Figure 1} a. They are characterized by the strong \ion{\rm Fe} {\sc \rm III} $\lambda$4404 \& $\lambda$5129 ~absorptions, visible \ion{Si}{III} $\lambda$4564, weak \ion{S}{II} W and \ion{Si}{II} $\lambda$6355 and lacking absorption features of \ion{\rm Ca} {\sc \rm II} H \& K and \ion{\rm Ca} {\sc \rm II} IR triplet before maximum optical light \cite{Ruiz_Lapuente92,Jeffery92}. 
The early-phase light curves of SN~2018evt are comparable to those of SN~1991T as shown in {\bf Extended Data Figure 1} b. The power-law fit of the earliest light curve of SN\,2018evt ($\lesssim-10$ days) suggests a rise time $t_{r}=18.76\pm0.24$ days, which is consistent with that of SN~1991T/1999aa-like events \cite{Ganeshalingam11_trise}.
The inset of {\bf Extended Data Figure 1} b shows the early-phase $B-V$ color curve, which is also in general agreement with that of SN~1991T after correcting the host reddening of $E(B-V) < $0.32 mag, which has been estimated from the equivalent width of the Na I D lines \cite{Yang22}.
The presence of the H$\alpha$ line makes it an SNIa-CSM similar to SN~2002ic. 
The entire spectral sequence of SN\,2018evt directly resembles other well-observed SNe Ia-CSM events such as PTF11kx and SN\,2002ic (see {\bf Extended Data Figure 2}). The NIR spectrum of SN~2018evt at $\sim 324$ days after the maximum is similar to that of another SNIa-CSM 
candidate SN 2012ca with data at a comparable epoch ({\bf Extended Data Figure 3}).

\subsection{Results}
\subsubsection{Observations}
SN~2018evt shows the characteristic spectral features of a Type Ia-CSM SN at early times together with Balmer emission lines ({\bf Extended Data Figure 1}) and the typical long-duration optical/IR light curves at late phases ({\bf Extended Data Figure 4}), indicating a continuous interaction between the expanding ejecta and a radially-extending CSM. 
We observed SN~2018evt with the {\it Spitzer} \cite{Werner04} InfraRed Array Camera (IRAC) at 3.6\ $\mu$m ($CH1$) and 4.5\ $\mu$m ($CH2$) \cite{Fazio04} in the year 2019 ({\bf Extended Data Figure 5 a} and {\bf Extended Data Table 1}). Meanwhile, the area of the SN location was scanned by the NEOWISE (near-Earth object $+$ the Wide-field Infrared Survey Explorer) reactivation mission \cite{neowise14} at 3.4\ $\mu$m ($W1$) and 4.6\ $\mu$m ($W2$) from 2019 to 2021 \cite{neowise14} ({\bf Extended Data Figure 5 b} and {\bf Extended Data Table 1}). 
The MIR fluxes of SN\,2018evt exhibit an initial decline from $+149$ to $+310$ days relative to the estimated $B$--band maximum at MJD 58352, however, it is followed by an unprecedented re-brightening until the SN reached its peak luminosity in both $W1$ and $W2$ bands at around day $+674$ ({\bf Figure 1 } a).
This behavior is not only distinct from the steadily-fading light curves in optical bandpasses but also has not been seen in any previous SNe\,Ia-CSM in similar MIR filters (see {\bf Figure 1} and {\bf Extended Data Figure 4} b, c); it is, however, likely that this is due to the lack of adequate time coverage of the observations of the latter. 

The optical spectral sequence of SN\,2018evt spans days $+125$ to $+579$ and also reveals conspicuous temporal evolution of the asymmetric characteristic H$\alpha$ profile.  
We measure the equivalent width (EW) separately for the red and the blue wings of the H$\alpha$ (shown in {\bf Extended Data Figure 2}, {\bf Extended Data Table 2} and \cite{Yang22}), Pa$\beta$, and Br$\gamma$ profiles (see in {\bf Extended Data Figure 3}, and {\bf Extended Data Table 2}). The ratios of red-to-blue wing flux increase steadily from day $+125$ to $\sim +$310 but turn over and decrease afterward ({\bf Figure 1} b), in pace with the MIR flux evolution.
Meanwhile, the flux-weighted centroid velocity $\Delta V$ of the H$\alpha$ line (see Methods 1 on analysis of the spectroscopic behaviors of SN\,2018evt for details) evolves steadily from the blue-shifted side ($-400$ \kms) to the red-shifted side ($+300$ \kms) before day $+310$, and thereafter moves gradually back to the blue side ($-200$ \kms) (see, {\bf Figure 1} c). 
In addition, the evolution of the EW of the
\ion{\rm Ca}{\sc \rm II}
NIR triplet also exhibits a fall and rise, in concert with the evolution of the MIR flux and the H$\alpha$ line profile (see, e.g., {\bf Figure 1} d).

\subsubsection{Model}
The slowly declining luminosity in the optical and NIR ({\bf Figure 1} and {\bf Extended Data Figure 4}) and the broad, long-lived H$\alpha$ line (as shown in \cite{Yang22} and {\bf Extended Data Figures 2, 6}) dominating the late-time spectra of SN\,2018evt, both indicate a substantial amount of late-time emission would arise from kinetic energy from the ejecta-CSM interaction converted to radiation \cite{Kotak04}. Such an additional energy source leads to a much slower luminosity decline ({\bf Figure 1} a) as powered by the $^{56}$Co$\rightarrow ^{56}$Fe decay, i.e., $\approx$0.97 mag\,100\,day$^{-1}$. 
In such a context, a cold, dense shell (CDS) develops during the ejecta-CSM interaction, with the CDS being located at a region between the shocked CSM and the shocked ejecta \cite{Chugai04:02ic,Moriya13}.
This is the region where the SN ejecta and the CSM mix produce suitable conditions that allow the condensation of dust grains on short timescales \cite{Sarangi18:10jl,Szalai19,Sarangi22}.

Blackbody fitting of the Spectral Energy Distribution (SED) of SN\,2018evt over the optical-to-NIR wavelength range suggests a broad temperature range of around 6400 K to 7000 K during our observations. 
As shown in {\bf Figure 2}, the MIR flux excess is obvious and becomes progressively more dominant over time as the optical emission decreases. 
The MIR flux excess can be attributed to the thermal emission from dust at temperatures of 100-1000 K \cite{Fox10,Szalai19}. 
The photospheric radius $R_{\rm BB}^{\rm Opt}$ estimated from the blackbody fitting is shown in {\bf Figure 3}. After day $+141$, $R_{\rm BB}^{\rm Opt}$ decreases continuously with time ({\bf Figure 3}).
 This indicates a progressive deviation of the blackbody photosphere from the expanding CDS  (e.g., Figure 7 of \cite{Smith10:06gy}), allowing the CDS to cool to a lower temperature.

We explore various radial profiles of pre-existing CSM that may account for the time-variant excess of MIR flux due to the thermal emission of dust. An initial decrease before day $+310$ could be attributed to a single-shell IR echo or a prominent process of dust sublimation as the forward shock runs through. 
The subsequent brightening after day $+310$ would suggest that newly formed dust accounts for the later MIR emission, in either the post-shock regions of the CSM or the cooling ejecta. Assuming CSM dust density follows a power-law distribution $\rho_{\rm dust} \propto r^{-s}$, the plausible fit to the time-variant MIR flux excesses before day $+310$ ({\bf Figure 4}) by searching among a grid of parameters requires a power-law index $s=1.15$. Other free parameters include the total optical depth and the inner and outer radius of the CSM shell (see Methods 2). The shallower radial density profile ($s=1.15$)  implies a rather enhanced dust content at larger distances from the progenitor star. 

In the case of the steady dust mass loss $s=2$, a plausible fit can also be achieved by introducing two shells of pre-existing CSM dust before day $+310$, namely the double-shell model. As the forward shock propagates outwards, grain sublimation only progressively takes place within the inner shell at a distance of $ 2.2 \times 10^{16}$ cm, while the emitting dust grains in the outer shell, which is located at $ 6.0 \times 10^{17}$ cm from the SN remain unaffected early on (See {\bf Figures 3, and 5}). Because of the lack of early-time spectral coverage, we adopt an initial shock velocity $V_{s}\approx$10000\,km\,s$^{-1}$ before day $+120$ based on the value estimated for SN\,2002ic \cite{Deng04,Fox13-05gjMIR}. As evidenced by the decreasing FWHM width of the H$\alpha$ profile (see inset of {\bf Figure 3}), the forward shock expands into the inner shell of the CSM and decelerates. The progressive destruction of the inner shell dust grains leads to a continuously decreased emission in the MIR ({\bf Figure 4}). After $\sim +$310 days, the forward shock supersedes the outer bound of the inner shell and enters a relatively low-density zone between the two CSM shells. The MIR emission becomes increasingly dominated by the relatively constant thermal emission from the outer shell. Our modeling suggests a rather massive outer shell of $5.2 \times 10^{-2}$\,M$_{\odot}$ of dust and an inner shell of $3.2 \times 10^{-5}$\,M$_{\odot}$ of dust, corresponding to two episodes of elevated dust mass-loss of $2.1 \times 10^{-5}$\,M$_{\odot}$\,yr$^{-1}$ and $1.8 \times 10^{-7}$\,M$_{\odot}$\,yr$^{-1}$, respectively (see Methods 2).

Akin to the single-shell model, our double-shell model also suggests enhanced dust concentration at larger distances from the SN as compared to what can be expected from the density profile of the mass loss from a steady stellar wind.  The dust distribution inferred from the MIR flux excesses before day $+310$ can be modeled in terms of a double-shell which assumes a sudden change of the density profile of the dust or a single-shell model with a flatter radial profile, see {\bf Figure 4}.

Both the single-shell and double-shell models are compatible with the MIR flux excesses at day $\lesssim +310$, but they can not fit the MIR flux excesses at day $> +310$. 
After day $+310$, the rebrightening of SN\,2018evt in the MIR demands significant contribution by additional emission sources, which can be well-attributed to the emergence of warm dust in regions behind the forward shock. As shown in {\bf Figure 3}, the blackbody radius $R_{\rm BB}^{\rm MIR}$ of the newly-formed dust content fitted to the SED after day $+$310, $R_{\rm BB}^{\rm MIR}$ increases monotonically and remains within the shock radius $R_{\rm s}$.

The inferred mass of the newly-formed dust increases over time follows a relation $M_{\rm d} \propto t^{4}$ and reaches $1.2\pm0.2\times 10^{-2}$\,M$_{\odot}$ by the time of our last observations at day +1041 (see {\bf Figure 6} and {\bf Extended Data Table 1}). The errors of the dust mass and temperatures are deduced using the Monte Carlo method via propagation of optical-to-NIR photometric errors into blackbody fits and the MIR photometric errors into the flux excess calculations. 
The dust sublimation time scale is extremely sensitive to the temperature close to the binding energy of the dust particles \cite{Wang:Wheeler:1987A}. Dust survival close to the shock is possible if the dust distribution is patchy or in an opaque disk in which the self-shielding of the dust particles is important \cite{Wang05}. Our double-shell model assumes that a substantial amount of dust may survive the initial UV/optical emission of the SN explosion out to the inferred inner CS dust shell radius of $2.2\times10^{16}$ cm, as shown in {\bf Figure 3}.

Moreover, our model with dust formation is also consistent with the time evolution of the observed colors in the optical. The colors can be modeled by including the absorption and scattering effects of the newly-formed dust ({\bf Extended Data Figure 7}). The increasing amount of dust after day $+310$ may contribute to the apparent blueward evolution of the $B-V$, $g-r$, and $g-i$ colors.
An increasing amount of scattered light is expected with more dust which leads to an excess flux in the $B$ and $g$ bands, as shown in {\bf Extended Data Figure 7}.
At even later epochs after day $\approx +500$, the SN also exhibits an accelerated fading in optical bandpasses, which is compatible with a change from the optically thin to optically thick regimes of the newly formed dust. Such a transition is similar to the dust-formation process observed in the ejecta of SN\,1987A \cite{Lucy89:87A}.

\subsection{Discussion}
The H$\alpha$ emission is powered by the interaction between the ejecta and the CSM \cite{Hamuy03:sn2002ic,Wang04}.
A thorough investigation of the time series of spectroscopy and spectropolarimetry within the first year of the SN explosion suggests a procedure that the SN ejecta expands into a dense torus of disk-like CSM \cite{Yang22}.
Such a configuration is in good agreement with the picture depicted by the spectroscopic and MIR flux evolutions span days $+125$ to $+1041$.
The SN ejecta running into a highly asymmetric disk-like CSM leads to a high-density torus inclined at an angle toward the observer. 
The early blueshift of the H$\alpha$ emission line is explained if the redshifted side of the shocked CSM is blocked by the photosphere, as shown schematically in {\bf Figure 5}. The redward shifts of the H$\alpha$ emission line (i.e., from days $+125$ to $+310$; see {\bf Figure 1}) are caused by the receding photosphere as the photosphere shrinks, as proposed for the CSM configuration in PTF11iqb \cite{Smith15:PTF11iqb} (their Figures 10 and 12). 
After day $+310$, warm dust grains start to coagulate in the CDS and gradually block the receding side of the H$\alpha$ line again, resulting in a blueward shift of the line profiles ({\bf Figure 5} c). 
After day $+674$, the $W1-W2$ color of SN\,2018evt becomes redder over time (see {\bf Figure 1} a), indicating a decrease in the temperature and the MIR emission of the newly-formed dust.

The presence of a highly asymmetric ejecta-CSM interaction zone is also supported by detailed spectropolarimetry of SN~2018evt which shows a wavelength-independent degree of polarization with non-evolving position angles that is characteristic of electron/dust scattering from a highly axisymmetric configuration \cite{Wang04,WangARAAdoi:10.1146/annurev.astro.46.060407.145139,Yang22}. 
Despite assuming spherical symmetry, both the single-shell and double-shell distribution of the CSM dust shell provides a satisfactory description of the SED evolution of SN\,2018evt spans days $+$149 to $+$310, in particular, the time-variant excess in MIR. By incorporating the geometric information obtained from spectropolarimetry \cite{Yang22}, the ejecta-CSM interaction process of SN\,2018evt before day $+310$ is illustrated by the schematic sketches (a) and (b) presented in {\bf Figure 5}.
In our double-shell model, the dust in the CS wind at the radius $\sim 2.2\times10^{16}$ cm may be distributed in a disk or torus instead. The destruction and formation of the dust would manifest qualitatively similar trends in the temporal evolution of the MIR excess.

The CSM masses derived from optical and optical-to-MIR luminosities in shock interaction regions are $\sim 0.2 - 4.5\ {\rm M}_\odot$ (see Methods 3 on the progenitor's mass loss for details; \cite{Yang22}), corresponding to mass loss rates of $\dot{M}=1\times 10^{-3}\ - \ 9\times 10^{-2}\ \rm M_\odot\ \rm yr^{-1}$. 
Such CSM masses estimated from the kinetic-to-radiation energy process across the shock front appears to be $\sim10^{5}$ larger compared to the amount of dust within the inner CSM shell ($3.2 \times 10^{-5}$\,M$_{\odot}$) that contributes most of the MIR excess before day $+$310 ({\bf Figure 4}). Thus a very low dust-to-gas mass ratio within the inner shell at a relatively smaller distance ($2.2\times10^{16}$\,cm) can be inferred, which may likely be caused by the prompt destruction of a substantial amount of grains in the inner shell by energetic particles from the SN \cite{Hoflich95:DDT}.
On the contrary, a gas-to-dust mass ratio on the order of 100 can be inferred in the more massive ($5.2\times 10^{-2}$\,M$_{\odot}$) and distant ($6.0\times10^{17}$\,cm) outer shell, which is consistent with what is anticipated in the interstellar medium \cite{Tricco17}. This probably means that the dust in the outer shell is much less affected by both the radiation field of the SN and the energetic particles from the shock interaction between the ejecta and the inner shell. Similar results as the outer shell can be derived by comparing the above CSM masses and the dust mass ($6.0\times 10^{-3}$\,M$_{\odot}$) located at $2.6\times10^{17}$\,cm in the single-shell model.

The mass loss of the progenitor before the explosion is in favor of either a thermonuclear explosion from a WD $+$ asymptotic giant branch (AGB) star system \cite{Hamuy03:sn2002ic,Inserra16}, or a core-degenerate system in which a WD merges with the core of a massive AGB star that triggers a thermonuclear explosion at the end of a common envelope phase or shortly after \cite{Soker13}. 
The mass loss is also consistent with a WD $+$ main sequence systems for the common envelope wind (CEW) model \cite{Meng17,Meng18}.
The progenitor systems are consistent with the measurements of the wind velocity $V_{\rm v}=91\pm58\ $\kms from the absorption minimum of the narrow P-Cygni profiles of H$\alpha$ line ({\bf Extended Data Figure 6} and that in \cite{Yang22}. 
Compared with the density profile of the dust mass loss from a steady stellar wind $s=2$, a flatter radial profile $s=1.15$ inferred in the single-shell model indicates enhanced dust concentration at increasing distances from the SN. The double-shell model also points to the same result that a higher dust mass-loss rate ($\dot{M}=2.1\times 10^{-5}\ \rm M_\odot\ \rm yr^{-1}$) of mass ejections is measured within the distant outer shell ($6.0\times10^{17}$\,cm) and a lower dust mass-loss rate ($\dot{M}=1.8\times 10^{-7}\ \rm M_\odot\ \rm yr^{-1}$) is measured within the close inner shell ($2.2\times10^{16}$\,cm). Both the single-shell and double-shell models suggest enhanced dust presence at larger distances from the progenitor star. This shallower radial density structure would result from a variable mass loss, which is likely to happen in the entire AGB evolution \cite{Hofner_Olofsson2018_massloss_AGB}.

The re-brightening in the MIR after day $+310$ can be modeled as a result of the formation of a substantial amount of warmer dust at late phases (see {\bf Figures 1, 2, and 4}), distributed in a prolate shell vertical to the CSM disk ({\bf Figure 5}). 
It also provides a natural explanation of the red-to-blue emission-wing ratio of H$\alpha$ due to uneven extinction by the newly-formed dust ({\bf Figure 1}). This behavior is also observed for the 
\ion{Ca}{II} NIR triplet which can be similarly explained. The rapid weakening of the \ion{\rm Ca}{\sc \rm II} NIR triplet may also indicate the depletion of calcium by dust formation. 
The proposed process of dust formation is corroborated by the time-evolution of the EW of the \ion{\rm Ca}{\sc \rm II} NIR triplet emission lines, which exhibits an increase after day $\sim +$310 (see, e.g., {\bf Figure 1} d).

{\bf Figure 6} shows the temporal evolution of the mass of the newly formed dust of SNIa-CSM 2018evt in our model, compared with other CC SNe. The estimated dust mass is highly dependent on the species and the size distribution of the dust grains. 
For graphite with a size of $0.3\ \mu$m  the dust mass grows rapidly following a power law of index 4 with the time after the explosion and reaches $\sim 1.2\pm0.2\times 10^{-2}$\ \Msun \ at the last epoch of the MIR observations at day $+1041$ ({\bf Figure 6}). For graphite or silicate of $0.05\ \mu$m, the dust mass is about three or five times higher than the value derived for the $0.3\ \mu$m graphite dust ({\bf Extended Data Table 1}), respectively. The temperature of the newly-formed dust is presented in the inset of {\bf Figure 5}. A monotonically decreasing temperature from 1000\,K to 500\,K between day $+434$ and day $+$1041 is likely to be mostly affected by the expansion cooling of the CDS region between the forward and the reverse shocks. It may also be regulated by various heating mechanisms, including radiative heating from the SN shock, collisional heating with the ambient warm gas, and the energy exchange between the gas and dust \cite{Sarangi18:10jl}.
As the SN ejecta expands and the dust-forming region in the CDS cools, dust grains may continue to coagulate. Depending on the duration of the time scale that the ejecta expands into the CSM, even orders of magnitude higher dust content may be produced during such a procedure.
A significant fraction of the unburned carbon in the ejecta, if not all, can be locked in the newly formed dust. As progressively deeper layers of the ejecta move into the CDS, we may also expect a massive amount of iron and silicate dust to form in such an environment. JWST can probe the signatures of such dust in the coming years. 
The estimated mass ($M_{\rm d}$) and the blackbody emission radii $R_{\rm BB}^{\rm MIR}$ of the newly formed dust masses are consistent with that seen in CC SNe at similar phases ( see {\bf Figures 6, 3}, and \cite{Sarangi18:10jl}), suggesting a rapid and efficient mechanism for dust production in these SNe. 

Finally, we remark that $\lesssim$1\% of Core-collapse SNe occur in elliptical galaxies in the local universe \cite{2022ApJ...927...10I}, dust production in thermonuclear explosion SNe Ia would be a major channel of dust enrichment in early-type galaxies.
SN Ia may also contribute to the dust budget in spiral galaxies \cite{Pipino11}. SN 2002ic was the first SNIa-CSM ever discovered and has a dwarf elliptical host \cite{Hamuy03:sn2002ic}. The weak H$\alpha$ of the host galaxy of SN\,2018evt  would also imply an overall less active star formation \cite{Jones09}.
Even though SNIa-CSM is a rare sub-class of thermonuclear SN, the unprecedented witness of such an intense production procedure of dust grains may shed light on the perceptions of dust formation in cosmic history (see Methods 4). 

\clearpage

%method
%\renewcommand\thefigure{S\arabic{figure}}
%\renewcommand\thefigure{\arabic{figure}}
\setcounter{secnumdepth}{0}
\section{Methods}
\subsection{1 Observations}
\subsubsection{Early-phase observations of SN~2018evt}

The early-phase observations of SN\,2018evt were conducted with the dual-channel optical/NIR camera ANDICam on the Cerro Tololo Inter-American Observatory (CTIO) 1.3-meter telescope. Two epochs of $BVRI-$band photometry were obtained on 2018-08-13 and 2018-08-17 before the SN was too close to the Sun.
ANDICam has an optical field of view (FOV) of $6'.3\times 6'.3$ ($0''.37$ pixel$^{-1}$) and a NIR FOV of $2'.4\times 2'.4$ ($0''.27$ pixel$^{-1}$).  
The extraction of the NIR-band photometry was not successful due to the lack of bright stars for astrometric calibration to combine the dithered images. The $BVRI$ point-spread function (PSF) photometry was performed on the optical images using PSFEx \cite{Bertin06} following the detailed prescriptions described by \cite{Wang20}.
The PSF photometry was calibrated to the standard Pan-STARRS catalog \cite{Tonry12,chambers16} of the brightest field star at (R.A., Decl.) = ($206.655661^{\circ}$,$-9.680946^{\circ}$) (J2000) with of $B=16.495\pm0.034$~mag, $V=15.812\pm0.012$~mag, $g=16.031\pm0.002$~mag, $r=15.603\pm0.002$~mag, and $i=15.461\pm 0.003$~mag. 
The $r-$ and $i-$band photometry of this field star has been converted to the standard Johnson $RI$ system \cite{Johnson1966} following the transforming equations provided by \cite{Jester05,Jordi06,Lupton05}.
The early optical light curves of SN\,2018evt are shown in {\bf Extended Data Figure 1} b. We also retrieve early-time photometry of SN\,2018evt using the ATLAS 
\cite{Tonry18,Smith20} forced photometry service in $c$ and $o$ bands, and the All-Sky Automated Survey for Supernovae  (ASAS-SN \cite{Shappee14,Kochanek17}) sky patrol interface. 
The background flux of the ASAS-SN data of SN\,2018evt has been estimated by the pre-explosion median flux recorded with the same aperture as used for the SN photometry, based on a total of 266 visits.
Both ATLAS and ASAS-SN photometry are also shown in {\bf Extended Data Figure 1} b. \\

\subsubsection{\textit{Spitzer} Observations}
SN~2018evt was observed (PI: Sijie Chen) with the {\it Spitzer} \cite{Werner04} Infrared Array Camera (IRAC) at 3.6~$\mu$m ($CH1$) and 4.5~$\mu$m ($CH2$) \cite{Fazio04} at days $+271$, $+286$, $+434$, and $+445$.  
We utilized the level 2 post-BCD (Basic Calibrated Data) images from the \textit{Spitzer} Heritage Archive (SHA), which were reduced by the {\it Spitzer} pipeline and resampled onto $0.6''$ pixels. Source detection and aperture photometry were performed on the images in {\bf Extended Data Figure 5} a without host subtraction using SExtractor \cite{sextractor}. We remark that flux difference is less than or similar to $10\%$ for the {\it Spitzer}/IRAC photometry with and without template subtraction in \cite{Szalai19}, which is well within the photometry uncertainty. We applied aperture corrections following the IRAC Data Handbook. 
The level 2 post-BCD images have been calibrated in an absolute surface-brightness unit of MJy/sr, which can be transformed into units of uJy/pixel$^2$ by a conversion factor of 8.4616 for the angular resolution of our IRAC images $0.6''$ pixels. The flux was converted to AB magnitude according to the definition $m_{AB}=-2.5log10(f)+8.9$, where $f$ is in units of Jy \cite{Fukugita96}. The AB magnitudes of SN 2018evt in $CH1$ and $CH2$ bands are listed in {\bf Extended Data Table 1}.

\subsubsection{NEOWISE observations}
SN~2018evt field was also observed by the NEOWISE reactivation mission in $W1$ ($3.4\ \mu$m) and $W2$ (4.6$\ \mu$m) bands since late 2013 as an extension of the WISE ALL-Sky Survey \cite{Wright10,Mainzer14}.
Using the online version of the NEOWISE Image Co-addition with 
Optional Resolution Enhancement (ICORE) 
\cite{Masci09,Masci13}, we retrieve the coadded NEOWISE images that centered at SN\,2018evt, with a FOV of $0.6^{\circ} \times 0.6^{\circ}$, and resampled to a pixel size of $1''.0$.
Given that SN~2018evt exploded in August 2018, we take the coadded image from January 2017 to January 2018 as the reference image for background subtraction, and generate the difference images for every single visit coadded image using the Saccadic Fast Fourier Transformation (SFFT) \cite{Hu22:SFFT}. {\bf Extended Data Figure 5} b shows the NEOWISE reference and difference images at the position centered on SN~2018evt. The time series of the differenced images clearly show the significant variations in the brightness of SN\,2018evt.
The signal was significant in January 2019. After a noticeable dimming in the next six months, a dramatic rebrightening is followed in 2020.

Aperture photometry was performed on the difference images using SExtractor \cite{sextractor} and calibrated to the profile-fit magnitudes in Vega system released in the ALLWISE Source Catalog. 
The photometric errors were measured on the corresponding variance images and corrected by a factor of 2.75, which gives the ratio of the input to output pixel scale (Section 13 of \cite{Masci13}). Such estimated photometric error $\sigma$ for each visit will be used if it is larger than the photon noise from direct photometry on the differenced images. The Vega magnitude of SN~2018evt was transformed into AB magnitude according to the magnitude offsets between the two magnitude systems \cite{Jarrett11}. The AB magnitudes of SN~2018evt in $W1$ and $W2$ bands are listed in {\bf Extended Data Table 1}. The MIR-band light curves of SN\,2018evt are shown in {\bf Extended Data Figure 4} b, c, together with other SNeIa-CSM, including SNe~2002ic, 2005gj \cite{Fox13-05gjMIR}, PTF11kx \cite{Graham17}, 2012ca \cite{Inserra14}, 2013dn \cite{Szalai19,Szalai21}, and 2020eyj \cite{Kool23}.

\subsubsection{Optical Photometry at Las Cumbres Observatory}

Extensive $BVgri$ photometry spanning $+$124 to $+$664 days was obtained with the Sinistro cameras on the Las Cumbres Observatory of the 1-m telescope, a global network for SN observations.
Images were bias subtracted and flat-field corrected using the BANZAI automatic pipeline. 
The background template was then subtracted from the pre-processed images adopting the SFFT algorithm \cite{Hu22:SFFT}. Finally, Point Spread Function (PSF) photometry has been performed on differenced images using \texttt{ALLFRAME} \cite{allframe}.
We remark that the light curves of SN\,2018evt before day $+365$ were achieved without subtracting any background template and reported in \cite{Yang22}. Comparisons between their direct photometry and our template-subtracted photometry obtained at similar phases suggest a good agreement. In particular, the systematic magnitude differences in $BVgri$ yield $-0.03\pm0.05$, $-0.03\pm0.05$, $0.02\pm0.05$, $-0.06\pm0.04$, and $-0.14\pm0.06$, respectively.

 Templates for our $gri-$band exposures were directly obtained using Panoramic Survey Telescope and Rapid Response System (Pan-STARRS) 
 cutout images for $gri$ bands. $B-$ and $V-$band templates were constructed using Pan-STARRS $gr$ images, with the formula $B=g+w\times (g-r)$, $V=g-w\times(g-r)$. 
 The parameter $w$ has been achieved by minimizing the global residual flux computed based on all field stars, which produces the cleanest subtraction.
 Thus the best coefficients obtained are $w= 0.3$ for $B$, $w=0.5$ for $V$, respectively.

Zero-point calibration was conducted using local field stars by calculating $3\sigma$ clipped median of the differences between instrumental magnitudes and standard Pan-STARRS Catalog \cite{Tonry12,chambers16} for $gri$ and AAVSO (American Association of Variable Star Observers) Photometric All-Sky Survey (APASS) data release 9 (DR9) catalog \cite{Henden16} for BV (using only stars with magnitude within 10-18 mag and photometric errors $\sigma < 0.1$~mag). The $BVgri-$band light curves of SN\,2018evt are shown in {\bf Extended Data Figure 4} a. 

\subsubsection{Optical Photometry with XLST \& LJT}

Optical photometric observations of SN\,2018evt were also conducted with the 60/90-cm XingLong Schmidt telescope (XLST) of National Astronomical Observatories of China (NAOC) under a long-term Tsinghua University-NAOC Transient Survey (TNTS) \cite{Zhang15}, and the Yunnan Faint Object Spectrograph and Camera (YFOSC) \cite{Zhang14} mounted on the 2.4-m LiJiang telescope (LJT) at the Yunnan Astronomical Observatories. 
SN2018evt was observed in the imaging mode of YFOSC.
Images obtained by the XLST and LJT were processed using an automatic custom pipeline based on the Image Reduction and Analysis Facility (IRAF). 
The pipeline reduction follows standard procedures including bias and flat-field corrections, astrometric registration, template subtraction, PSF photometry.
The $BVgri$ photometry is also shown in {\bf Extended Data Figure 4} a. 
\\
\subsubsection{Optical Spectroscopy}
 
We also obtained 17 optical spectra of SN\,2018evt. A log of the spectroscopic observations is presented in {\bf Extended Data Table 2}.
Six spectra were taken with the 3.6-m ESO Faint Object Spectrograph and Camera v.2 (EFOSC2) \cite{Buzzoni84} mounted on New Technology Telescope (NTT) at La Silla Observatory during the extended-Public ESO Spectroscopic Survey
for Transient Objects (ePESSTO) \cite{Smartt15}.
The observations were carried out under ESO programs 199.D-0143 (PI: Smartt) and 1103.D-0328, 106.216C ( PI: Inserra). Four spectra were taken with the YFOSC/LJT in the long-slit spectroscopic mode, and three were taken with the Beijing Faint Object Spectrograph and Camera (BFOSC) \cite{Fan16} mounted on the 2.16-m Xinglong telescope (XLT).
One spectrum was obtained with the Wide Field Spectrograph (WiFeS) mounted on the ANU (Australian National University) 2.3-m telescope at the Siding Spring Observatory \cite{Dopita07}. 
Three additional spectra obtained at days $+$490, $+$516, and $+$531 were acquired with the Folded Low Order whYte-pupil Double-dispersed Spectrograph (FLOYDS \cite{Sand11}) mounted on the 2.0-m telescope at Las Cumbres Observatory({\bf Extended Data Table 2}). 
The twin robotic FLOYDS spectrographs are mounted on the Faulkes Telescope South (FTS) at Siding Spring Observatory and on the Faulkes Telescope North (FTN) at Haleakala. 
Apart from the three late-time spectra mentioned above, another 12 spectra ({\bf Extended Data Table 2}) were obtained with the same telescope spanning from days $+125$ to $+365$ days \cite{Yang22} and were also included in this paper to measure the H$\alpha$ line profile (e.g., red-to-blue emission-wing ratio and flux-weighted centroid velocity $\Delta V$). 
The photometry of SN\,2018evt obtained with the global network of 1-m telescopes and the 12 epochs FLOYDS spectroscopy before day $+$365 have been published in \cite{Yang22}, which focuses on the early ejecta-CSM interaction and the spectropolarimetric properties of SN~2018evt.

All optical spectra were reduced using standard IRAF routines. 
Flux calibration of the spectra was carried out using spectrophotometric standard stars observed at similar airmass on the same night. The spectra were further corrected for atmospheric extinction using the extinction curves of local observatories.

\subsubsection{NIR Spectroscopy}
This paper includes eight NIR spectra (see {\bf Extended Data Figure 3} and {\bf Extended Data Table 2}). Four NIR spectra were obtained with the Medium-Resolution 0.8-5.5 Micron Spectrograph and Imager on the 3.0-m NASA Infrared Telescope Facility (IRTF) on Mauna Kea, named SpeX \cite{Rayner03}. Two NIR spectra were acquired with the Folded port InfraRed Echellette (FIRE) spectrograph \cite{Simcoe13} on the 6.5-m Magellan Baade telescope. Another two spectra were obtained with the Gemini near-infrared spectrograph (GNIRS) \cite{Elias06:GNIRS} on the 8.2-m Gemini North telescope.
The SpeX, FIRE, and GNIRS spectra were reduced with the IDL codes, \texttt{Spextool} \cite{Cushing04}, {\texttt firehose} \cite{Simcoe13}, and the \texttt{XDGNIRS} pipeline \cite{Hsiao15,Hsiao19}, respectively.

\subsubsection{Analysis of the spectroscopic behaviors of SN\,2018evt}

All spectra were corrected for the redshift $z=0.02523$ of the host galaxy \cite{Yang22}, and the extinction from the Milky Way  $E(B-V)=0.05$ mag \cite{SF2011}. Three spectra lines were normalized with a pseudo-continuum by linear fitting to the spectra ranges [6250, 6350] \AA~and [6700, 6800] \AA~for H$\alpha$, [12200, 12500] \AA~and [13200, 13500] \AA~for Pa$\beta$,  [21100, 21400] \AA~and [21900, 22200] \AA~for Br$\gamma$. Thus H$\alpha$, Pa$\beta$, and Br$\gamma$ are located at [6350, 6700] \AA, [12500, 13200] \AA, and [21400, 21900] \AA, respectively. 
All spectra were scaled to match the photometry in the optical bandpasses at corresponding phases and further used to measure the H$\alpha$ luminosity ({\bf Extended Data Table 2}) and EW ({\bf Figure 1} d). 
For each flux spectrum, following an approach similar to the analysis of \cite{Yang22}, we fit a double-component Gaussian function to the H$\alpha$ profile to decompose it into a broad and an intermediate component. We found that the center of the intermediate Gaussian component, which has a typical FWHM width of $\sim$2000\,km\,s$^{-1}$, shows only moderate shift over time until the last epoch of spectroscopy at day $+579$. Such behavior is in overall good agreement with the analysis based on the spectra obtained before day $+365$ by \cite{Yang22}. 
The determination of the center of the intermediate Gaussian component also allows us to compare the blue and red wings of several major emission lines. In particular, we present the red-to-blue EW ratios measured between the red and the blue wings for H$\alpha$, Pa$\beta$, and Br$\gamma$ features in {\bf Figure 1} b. Note that the determination of the center of the Pa$\beta$ and the Br$\gamma$ lines was carried out based on a single-component Gaussian fit due to the relatively low S/N. The EW ratios for the H$\alpha$ and the Pa$\beta$ lines were computed over a velocity range of $-$8000 to $+$8000\,km\,s$^{-1}$. A narrower velocity range of $-$3500 to $+$3500\,km\,s$^{-1}$ was used for the measurement of the Br$\gamma$ profile.

After correcting for the redshift of the host galaxy, we define the flux-weighted centroid velocity as $\Delta V=\frac{\lambda_{\rm peak} - \lambda_0}{\lambda_0} \times c$, where $c$ gives the speed of light, $\lambda_{0}$ represents the rest wavelength of the line center, i.e., $\lambda_0(\rm H\alpha)$=6563\AA. The flux-weighted peak wavelength, $\lambda_{\rm peak}$, is calculated as $\lambda_{\rm peak}=\frac{\int \lambda\ f d\lambda}{\int fd\lambda}$, where $\lambda$ denotes the wavelength of any spectral element over the emission profile. Such a quantity literary weights each spectral element by their flux $f$, thus providing a more robust trace of the bulk velocity of the line-emitting zone. {\bf Figure 1} c shows the $\Delta V$ derived for the H$\alpha$, Pa$\beta$, and Br$\gamma$ lines for SN\,2018evt. 
In {\bf Figure 1}, the uncertainties of EW ratio and centroid velocity were calculated through the Monte Carlo method, assuming that all spectra have 10\% flux uncertainty.

The day $+$307 WiFeS \cite{Dopita07} spectrum obtained with a higher spectral resolution (R$\approx$3000) presents a well-resolved narrow H$\alpha$ P Cygni profile (see {\bf Extended Data Figure 6}).
A two-component Gaussian fitting process suggests that the FWHM widths of the broad and the intermediate components are 5877$\pm$32\,km\,s$^{-1}$ and 1643$\pm$12\,km\,s$^{-1}$, respectively.
After subtracting the broad and the intermediate components from the day $+$307 WiFeS spectrum, we fit the residual spectrum with two separate Gaussian functions to better separate the narrow absorption and emission components of the P Cygni profile. 
We inferred a redshift $z=0.02561\pm0.00019$ by assuming the narrow emission component peaks at the rest wavelength of H$\alpha$ (see inset of {\bf Extended Data Figure 6}). The wind velocity, which is measured from the blue-shifted absorption minimum, gives $V_{w}=91\pm58$\,km\,s$^{-1}$. Our measurements are consistent with those reported by \cite{Yang22} within the uncertainties, for example, $z=0.02523\pm0.00015$ and $V_{w}=63\pm17$\,km\,s$^{-1}$. 
The redshift values derived in both studies are also consistent with those reported to the NASA/IPAC Extragalactic Database (NED) \cite{daCosta98,Theureau05}. Therefore, we used $z=0.02523$ and $V_{w}=63$\,km\,s$^{-1}$ \cite{Yang22} throughout the paper due to the smaller uncertainty.

\subsection{2 Blackbody fit and dust sublimation}
The Effective blackbody temperatures and radii $R_{\rm BB}^{\rm Opt}$ were estimated by fitting a black body (BB) curve to a time series of SED constructed from the optical ($BVgri$) and/or NIR ($JHK_{s}$) light curves of SN\,2018evt.
Optical photometry was obtained by the global network of the 1-m telescope at Las Cumbres Observatory, and NIR photometry was taken from \cite{Yang22}. The latter spans days $+$141 to $+$314, and was obtained with the Gamma-Ray burst Optical/Near-Infrared Detector (GROND) \cite{Greiner08} mounted on the 2.2-m MPG/ESO telescope, operated at the La Silla Observatory in Chile.
All SED were constructed after correcting for the $E(B-V)=0.05$\,mag Galactic extinction \cite{SF2011}.
We adopt a distance of 103.3\,Mpc for SN\,2018evt following the rationale provided in \cite{Yang22}. 
Owing to the lack of early-time data before day $+$141, we adopt the optical and NIR light curves of the well-sampled Type Ia-CSM SN\,2005gj \cite{Prieto07:sn2005gj} to generate the SED of SN\,2018evt during the missing phases, and thereafter to calculate the MIR emission of CSM dust through the absorption and re-emission processes.
Such an approximation is validated by the high similarity in pre-peak and around day 140 spectra, light curve shapes, and absolute brightness in optical and NIR bandpasses between SNe\,2018evt and 2005gj.
In detail, SNe~2018evt and 2005gj are both Type Ia-CSM objects (see {\bf Extended Data Figure 1} and Figure 7 in \cite{Aldering06:sn2005gj}). They share similar blackbody temperature and radius at day $\approx$ 140 based on their multiband photometry, as shown in {\bf Figure 3}, and Table 8 of \cite{Prieto07:sn2005gj}. Also, the peak fluxes of SN~2005gj are comparable with that of SN~1991T (Figure 7 of \cite{Prieto07:sn2005gj}),  whose early-phase spectrum and light curves match well with SN~2018evt \cite{class:18evt} as shown in {\bf Extended Data Figure 1} (see \cite{Filippenko92:91t, Mazzali95:91t,Lira98,Phillips22} for SN\,1991T).

For SNe whose late-time emission is mostly dominated by strong ejecta-CSM interaction, their 
effective $R_{\rm BB}^{\rm Opt}$ is expected to coincide with the radius of a thin CDS \cite{Chugai01,Chugai04:02ic} located between the shocked CSM and the shocked ejecta.
The $R_{\rm BB}^{\rm Opt}$ of SN\,2006gy reaches its maximum value at day $\sim +$115 (see Figure 7 of \cite{Smith10:06gy}), while the expansion of its CDS continues as indicated by a rather constant FWHM width of the H$\alpha$, which traces the expansion velocity of the CDS.
Similar behavior is also seen in the Type IIn SN~2006tf (see Table 3 and Figure 15 of \cite{Smith08:06tf}). 
Therefore, we suggest the black-body-fitted $R_{\rm BB}^{\rm Opt}$ does not trace the emitting radius of the CDS $R_{\rm CDS}$. The latter can be represented by introducing a dilution factor $\zeta$, which cannot exceed unity and decreases over time \cite{Smith10}. The true emitting radius of the CDS is given by $R_{\rm CDS}$= $R_{\rm BB}^{\rm Opt}/ \sqrt{\zeta}$ \cite{Smith08:06tf,Smith10}.

{\bf Figure 3} shows the $R_{\rm BB}^{\rm Opt}$ of SN~2018evt through blackbody fit to the optical-to-NIR photometry. At day $+141$, we measure $R_{\rm BB}^{\rm Opt}=3 \times 10^{15}$~cm. After day $+141$, $R_{\rm BB}^{\rm Opt}$ is decreasing in {\bf Figure 3}, indicating that $R_{\rm CDS}$ has departed from the corresponding blackbody radius at day $+141$ ($\zeta \lesssim 1$). We approximately adopted the $R_{\rm BB}^{\rm Opt}=3 \times 10^{15}$~cm as the lower limit of the expanding CDS radius at day $+141$. 

For SN ejecta whose radial density profile follows an inverse power law distribution, $\rho_{\rm ejecta} \propto r^{-n}$, the shock radius is given by the equation (1) of \cite{Fransson14}: 

\begin{equation}
    R_{\rm s}=9.47\times 10^{15}\frac{(n-2)}{(n-3)}(\frac{V_{\rm s}}{3000\ \rm km s^{-1} })(\frac{t}{\rm years})^{(n-3)/(n-2)} \ \rm cm
\end{equation}

In the literature, the shock velocity $V_{\rm s}$ is 
often approximated by the velocity corresponding to the FWHM width of the H$\alpha$ emission line \cite{Kiewe12,Taddia13,Kokubo:2019ApJ...872..135K}. Assuming a typical shock velocity of 5000 - 10000~\kms \cite{Deng04,Chugai17:06gy,Chugai18}, which is consistent with the H$\alpha$ velocity width measured during our spectroscopic observing campaign on SN\,2018evt between days $+$125 and $+$546, see inset of {\bf Figure 3}. By adopting a $n=$8.5 ejecta density profile estimated for SN\,2002ic \cite{WoodVasey04}, we estimate a forward shock radius $R_{\rm s} = 2.8 \times 10^{16}$\,cm at day $+$310.

This shock radius $R_{\rm s}$ is less than the dust evaporation radii of $\sim 4 - 9 \times 10^{16}$ cm  for silicate or graphite dust \cite{Fransson14} assuming the SN luminosity to be $L_{\rm bol}=10^{43}\ \rm erg ~s^{-1}$, which falls in between the maximum for SN~2018evt ($10^{42.8}\ \rm erg ~s^{-1}$) and the peak luminosity of SN~2005gj ($10^{43.7}\ \rm erg ~s^{-1}$ in Table 8 of \cite{Prieto07:sn2005gj}). This suggests that the pre-existing CSM dust in the single-shell model at $2.6\times10^{17}$ cm and in the outer shell of our double-shell model at a distance $6.0\times10^{17}$ cm are unlikely to be sublimated by the SN radiation as the dust temperature at $2.6\times10^{17}$\,cm can only be heated up to a temperature of about 970 K, which is lower than the evaporation temperature of 1500 K for silicate and 1900 K for graphite \cite{Fransson14,Gall14,Dwek21:10jl}. 
The shock radius $R_{\rm s}$ is comparable to the inner radius of the inner CSM dust shell in our double-shell model ($ 2.2\times 10^{16}$ cm), indicating 
the dust grains within the inner CSM shell are likely to be destroyed by the forward shock, 
if they survived the initial pulse of the electromagnetic radiation of the SN explosion due to a patchy dust distribution or in an opaque disk \cite{Wang05}.

\noindent
\textbf{MIR flux excess}

The MIR flux excess 
compared to the best-fit black-body SED for different epochs of observations is shown in {\bf Figure 2}. Two-epochs of MIR observations were acquired with {\it Spitzer} $CH1$ ($3.6\ \mu$m) and $CH2$ ($4.5\ \mu$m) at days $+271$ and $+445$. The MIR observations at days $+286$ and $+434$ were not presented as they are  nearly identical 
to the results for days +271 and +445 observations respectively ({see \bf Extended Data Figure 4, and Extended Data Table 1}). Six more epochs of observations were acquired with NEOWISE $W1$ ($3.4\ \mu$m) and $W2$ ($4.6\ \mu$m) at days $+149$, $+310$, $+517$, $+674$, $+881$, and $+1041$.

MIR emission excess typically suggests the presence of warm dust. 
The MIR filters used to observe SN\,2018evt provide a rather complete wavelength coverage of observations spanning the peak of the thermal spectral energy distribution from dust with temperature spanning $100\le T_{\rm d} \le 1000$~K \cite{Fox10,Fox11}. 
The MIR emission excess has been 
explained by the formation of new dust grains in a handful of CC SNe, both in the ejecta in-situ \cite{Rho08, Barlow10,Matsuura11,Gomez12a} and in the interactions between the ejecta and the CSM (e.g., \cite{Rho08, Smith09:05ip,Gall14}). 
Alternatively, MIR emission excess could originate from the thermal infrared radiation emitted by dust particles that were present in a CSM before the SN event. 
Such primordial dust grains may have formed in the expanding matter blown from the red giant stars or AGB stars \cite{Hamuy03:sn2002ic,Smith14:review}. 
In addition to the thermal radiation of the pre-existing CSM grains, our models also include the emission from any newly-formed dust to account for the extreme MIR rebrightening of SN\,2018evt after day $+$310.

\textbf{Modeling to the emissions of SN\,2018evt: a model with one or two primordial CSM shells and new dust formed in the CDS region}

Dust particles in the CSM absorb some of the UV-optical photons radiated 
during the explosion of the SN and its ejecta interaction with the CSM and re-emit the flux in the IR bands, producing an IR echo \cite{Dwek83,Draine_Lee1984,Laor_Draine1993,Weingartner_Draine2001,Dwek85,Maeda15}. Such an IR echo can be used to constrain the CSM dust properties around the SN such as their distribution, mass, and composition.
IR echo models for spherically symmetric CSM shells have been developed to account for the thermal emission from pre-existing CSM dust, which provides a plausible explanation for the late-time excess in the observed IR light curves Type Ia \cite{wang08-echo,Maeda15,Wang19} and Type II SNe \cite{Dwek83,Dwek85}. 
The time evolution of the IR echo is related to the ultraviolet and optical light curves of the SNe. 

At any given time, a distant observer will see the IR echo located within an ellipsoid, with the SN and the observer lying at its two foci. Such an ellipsoid traces an iso-travel-time surface of the light emitted by the SN, which expands over time. The position of any point within the ellipsoid can be expressed as ($r$, $\theta$), where $r$ denotes the distance from the point to the SN, and $\theta$ represents the scattering angle.
For dust particles of radius $a$ located at $(r,\theta)$, and an SN located at a distance $D$ from the observer, the total flux emitted by the IR echo at time $t$ gives
\begin{equation}
   F_v(t)=\frac{a^2}{D^2} \int_{R_{\rm in}}^{R_{\rm out}} n_d(r) \pi B_v[T_{\rm d}(r,\theta,t)] Q_v \,d^3r 
\end{equation}
where $R_{\rm in}$ and $R_{\rm out}$ are the radii of the inner and outer dust shell, respectively, $n_{\rm d}$ is the number density of the dust particles, the Plank function $B_v$ at frequency $\nu$ is determined by the dust temperature $T_{\rm d}(r,\theta,t)$, 
which can be estimated from the SN luminosity. 
$Q_{\nu}$ denotes the absorption and emission efficiency of the dust grains.  
The light curve of the emitting IR echo from dust distributed in a shell shows a plateau lasting for a period of 2$R_{\rm in}/c$ (where $c$ is light speed) and followed by a decline for a period of time that is related to the radial extent of the shell. 
Such behavior is similar to that reported for SN\,2005ip \cite{Smith09:05ip} and several other SNe that show strong ejecta-CSM interaction (e.g., \cite{Dwek83,Dwek85,Maeda15,Dwek21:10jl}). \\

Assuming the CSM dust density around SN\,2018evt follows an inverse power-law distribution $\rho_{\rm dust} \propto r^{-s}$, the MIR flux emitted by the CSM dust within a single shell can be derived from equation (2).
The single-shell model is assumed to be spherically symmetric and described by four parameters namely inner and outer radii of the shell $R_{\rm in}$, $R_{\rm out}$, the optical depth in $B$ band  $\tau_B$, and the power-law index $s$. We initially set $R_{\rm in}$ in the range of 50 ld $ <R_{\rm in}< $ 150 ld (ld for the light day), in order to make sure the thermal radiation of the CSM dust declines between 100 and 300 days. We run the single-shell model in tens of thousands of grid points based on the four parameters ($\tau_B$, $R_{\rm in}$, $R_{\rm width}$, $s$) and obtain a group parameter ( 0.07, 100 ld, 80 ld, 1.15) to well fit the data (see the green lines in {\bf Figure 4}), where $R_{\rm width}=R_{\rm out}-R_{\rm in}$. A flatter radial profile of the CSM dust density was inferred from the single-shell model due to the smaller $s$, compared with the value for the steady-wind mass loss of the progenitor system ($s=2$). This suggests an increased dust concentration at increasing distances from the SN as compared to what can be expected from the density profile of the mass loss from a steady stellar wind. The mass of the CSM dust within the single shell is derived to be $6.0\times 10^{-3}\ \rm M_{\odot}$.\\

In the case of the steady-wind mass loss $s = 2$, a plausible fit can also be achieved by introducing two shells of pre-existing CSM dust before day $+310$. 
The inner shell predicted by our double-shell model was caught by the forward shock at day $\sim +$200 (see, e.g., {\bf Figure 3}). Dust grains within the inner shell are thus gradually destroyed as the shock runs. The expansion velocity of the shock was assumed to be $V_{\rm s}\approx$10000\,km\,s$^{-1}$ due to the lack of observations before day $+$120, followed by a continuous deceleration as traced by the FWHM width of the broad H$\alpha$ as shown by the inset of {\bf Figure 3}. 
Before day $+1041$, the outer shell of the CSM, which emerges at $6.0 \times 10^{17}$\,cm, remains unaffected by the forward shock.
Our double-shell CSM model also provides a satisfactory fit to the monotonically decreasing MIR flux curves before day $+310$. 
The SED fits to the $BVgriJHK_{s}$ and MIR$-$band photometry are illustrated in {\bf Figure 2}. The IR echo light curves of the two shells are shown in {\bf Figure 4}. 
The best-fit parameters are ($\tau_B$, $R_{\rm in}$, $R_{\rm width}$)=(0.07, 8.5 ld, 2.2 ld) for the inner shell and (0.17, 230 ld, 30 ld) for the outer shell. 
The total optical depth of the pre-existing dust shells in $B-$band is $\tau_B=0.24$, corresponding to a $V-$band extinction of $A_{V}=0.26$~mag. 
We also remark that such an integrated extinction is consistent with the value estimated from the Na I D lines \cite{Yang22}. 
By assuming the CSM shells were built up by multiple epochs of pre-explosion eruptions, the derived dust mass-loss rates of the mass ejections that form the
inner and outer dust shells are $1.8\times 10^{-7}\ \rm M_{\odot} \rm yr^{-1}$ and  $2.1\times 10^{-5}\ \rm M_{\odot} \rm yr^{-1}$, yielding total dust masses of $3.2\times 10^{-5}\ \rm M_{\odot}$ and $5.2\times 10^{-2}\ \rm M_{\odot}$, respectively. 

However, newly-formed dust is required to explain the substantial elevation of the MIR flux excess at day $> +310$. The fit results of the MIR excess of SN\,2018evt are also shown as red-dotted curves in {\bf Figure 4}. 
Our fit result is achieved by assuming the newly-formed dust is composed of graphite grains of radius $a=0.3\,\mu$m. In {\bf Figure 6}, we also present the mass of the newly-formed dust as a function of time for $a=0.05\,\mu$m graphite and silicate grains. 
Dust masses estimated for other well-sampled SNe that exhibit ejecta-CSM interactions are also presented for comparison, including the Type IIP SNe 2004et (diamonds) \cite{Fabbri11:04et} and 1987A (blue upward triangles) \cite{Spyromilio88,Ercolano07,Bevan16}, and IIn SNe 2005ip (crosses) \cite{Bevan19:05ip:ejecta,Stritzinger12}, 2006jd (squares) \cite{Stritzinger12}, and 2010jl (green downward triangles) \cite{Maeda13,Gall14}. As shown in {\bf Figure 6}, the amount of dust formed by SN\,2018evt is equivalent to those formed in CC SNe.

We also remark that at day $< + 310$, the MIR flux excess measured in band 2 (CH2 or W2) is higher than or comparable to that in band 1 (CH1 or W1), indicating a higher dust emission efficiency towards longer wavelengths. 
This manner is compatible with the large ($a=1.0\,\mu$m) graphite dust particles in the primordial CSM shells suggested by our single-shell and double-shell models to the time-evolution of the MIR excess at day $< +$310.

The $a=0.3\,\mu$m graphite dust model provides satisfactory fits to both the MIR photometry in bands 1 and 2 at day $>+310$ (see {\bf Figures 2 and 4}). The indicated best-fitting radius of the newly-formed dust grains also falls within the 0.01 to 1\,$\mu$m range of the typical size of the graphite dust grains (e.g., \cite{Fox11,Dwek21:10jl}). However, the species of the newly-formed dust grains may still not be inferred based on our observations as no spectral signatures of CO overtone bands at 2.3-2.5 $\ \mu$m were seen from our NIR spectra shown in {\bf Extended Data Figure 3}, see also SN~2017eaw \cite{Rho2018:17eaw}.
Additionally, we are not aware of any observation of SN\,2018evt conducted at 9\,$\mu$m, which may discriminate the silicate and graphite dust models \cite{Draine_Lee1984,Fox11}.
Therefore, we also present the results computed for $a=0.05\,\mu$m silicate and graphite dust grains in {\bf Figure 6} and {\bf Extended Data Table 1}.

\subsection{3 The progenitor's mass loss}
\label{massloss}

Before the SN explosion, the progenitor mass-loss rate $\dot{M}$ can be associated with the bolometric luminosity via a 
factor $\epsilon$, which denotes the kinetic-to-radiation energy conversion efficiency. Assuming a steady stellar wind CSM (s=2 in $\rho_{\rm csm} \propto r^{-s}$ \cite{Moriya13}), the bolometric luminosity $L_{\rm bol}$ can be written as:

\begin{equation}
    L_{\rm bol}=\epsilon\frac{dE_{\rm kin}}{dt}=\frac{1}{2}\epsilon\frac{\dot{M}}{V_{\rm w}}V_{\rm s}^3
\end{equation}
where $E_{\rm Kin}$ represents the kinetic energy of the thin shocked shell. 
The efficiency factor $\epsilon$ is often assumed between 0.1 and 0.5 \cite{Fox11,Moriya13,Moriya14,Ofek14}. We adopted $\epsilon=0.3$ and $V_{\rm s}=2000$\ \kms, the latter is consistent with the typical FWHM velocity of the intermediate H$\alpha$ component measured over our spectroscopic campaign on SN\,2018evt. 
The wind velocity blown from the progenitor was taken from the P-Cygni feature reported in \cite{Yang22}, $V_{\rm w}=63$~\kms.
A similar velocity was only observed in the unshocked CSM of PTF11kx \cite{Dilday12} ($V_{\rm w} \sim$ 65 \kms), an SN~1999aa-like SN, which 
exhibits multiple CSM components but displays no signature of the ejecta-CSM interactions based on the early-time observations \cite{Dilday12,Silverman13:PTF11kx}. 

A sudden decrease in the optical light curves of SN\,2018evt can be seen at day $\approx +$530, indicating the formation of new dust grains in the CDS (e.g., {\bf Figure 1} and {\bf Extended Data Figure 4}).
Following the prescription in \cite{Yang22}, we approximate the optical bolometric luminosity ($L_{\rm Opt}$) of SN\,2018evt by integrating its SED at day $+$530 over the optical wavelength range (3870$-$9000\,\AA). The day $+$530 SED was obtained by warping the day $+$264 flux spectrum to match the $BVgri-$band photometry at day $+$530. Therefore, the estimated 
$L_{\rm Opt}=5.2 \times 10^{41}$~erg s$^{-1}$ at day $+530$ 
yields a mass loss rate of 

\begin{equation}
    \dot{M} \approx 0.04 \ \rm M_{\odot}\, yr^{-1} (\frac{L_{\rm bol}}{5.2\times 10^{41}  \ \rm erg\, s^{-1}}) (\frac{V_w}{63 \ \rm km\, s^{-1}}) (\frac{0.3}{\epsilon}) (\frac{2000  \ \rm km\, s^{-1}}{V_{\rm s}})^3 
\end{equation}

The mass of shocked CSM around SN~2018evt can be estimated 
by multiplying the mass-loss rate to the duration of the shock propagation ($t_{\rm duration}$) as approximated by the phase of the measurement $t_{\rm duration}=530$~days, which can be expressed as

\begin{equation}
    M_{\rm shocked~CSM}=\frac{V_{\rm s}}{V_{\rm w}} \dot{M} \times t_{\rm duration}  \\
    \approx 2.0\ {\rm M}_{\odot} (\frac{L_{\rm bol}}{5.2\times 10^{41}  \ \rm erg\, s^{-1}}) (\frac{0.3}{\epsilon}) (\frac{2000  \ \rm km\,s^{-1}}{V_{\rm s}})^2 \times (\frac{t_{\rm duration}}{530 \ \rm days})
\end{equation}

At such late phases of SN\,2018evt, the dominant radiation source in the IR can be well attributed to the thermal emission of newly-formed dust (see {\bf Figure 2}). 
Thanks to the MIR observations at day $+$517, a phase comparable to $+530$, we estimate the optical-to-MIR pseudo-bolometric luminosity $L_{\rm Opt+MIR}$ of SN\,2018evt by integrating the SED over a wavelength range of 3870\,\AA--5\,$\mu$m (see {\bf Figure 2}). The computed $L_{\rm Opt+MIR} = 1.2 \times 10^{42}$\,erg\,s$^{-1}$ at day $+517$ indicates a mass-loss rate $\dot{M}\approx$0.09\,M$_{\odot}$\,yr$^{-1}$. Therefore, the corresponding mass of the shocked CSM can be estimated to be $M_{\rm shocked\,CSM}\approx$4.5\,M$_{\odot}$.

At day $+517$, 
adopting a shock velocity $V_{\rm s}=$6000\,km\,s$^{-1}$ estimated by the FWHM width of the broad H$\alpha$ component (see \cite{Yang22} and {\bf Figure 3}), and an optical bolometric luminosity $L_{\rm Opt} = 5.2 \times 10^{41}$\,erg\,s$^{-1}$, following Equations (4) and (5), the corresponding mass-loss rate and the mass of the shocked CSM yield $\dot{M}\approx$0.001\,M$_{\odot}$\,yr$^{-1}$ and $M_{\rm shocked\,CSM}\approx$0.2\,M$_{\odot}$, respectively. 
If we include the emission in the MIR by adopting $L_{\rm Opt+MIR} = 1.2 \times 10^{42}$\,erg\,s$^{-1}$, the corresponding $\dot{M}$ and $M_{\rm shocked\,CSM}$ can be estimated to be 0.003\,$M_{\odot}$\,yr$^{-1}$ and 0.5\,$M_{\odot}$, respectively.

\ \par

In the literature, the luminosity of H$\alpha$ line $L_{\rm H\alpha}$ serves as a good indicator of the bolometric luminosity $L_{\rm bol}$ as $L_{\rm H\alpha}$ has found to be proportional to $L_{\rm bol}$ \cite{Salamanca98,Kiewe12,Taddia13,deJaeger15,Kokubo:2019ApJ...872..135K}. $L_{\rm H\alpha}$ can be expressed as

\begin{equation}
    L_{\rm H\alpha}=\frac{1}{2}\epsilon_{\rm H_{\alpha}} \frac{\dot{M}}{V_{\rm w}}V_{\rm s}^3
\end{equation}

where $\epsilon_{\rm H_{\alpha}}$ denotes the efficiency of the conversion of the dissipated kinetic energy into H$_{\alpha}$ luminosity in the shock wave. For SN~2018evt at day $\sim +530$, 
we measured $L_{\rm H\alpha}=3.4 \times 10^{40}\rm erg\,s ^{-1}$ ({\bf Extended Data Table 2}). Thus we can get

\begin{equation}
    \frac{\epsilon_{\rm H_{\alpha}}}{\epsilon} = \frac{L_{\rm H\alpha}}{L_{\rm Opt+MIR}} \\
    \approx 0.03 \times (\frac{L_{\rm H_{\alpha}}}{3.4\times 10^{40}  \rm erg s^{-1}}) (\frac{L_{\rm Opt+MIR}}{1.2\times 10^{42}  \rm erg s^{-1}})^{-1}
\end{equation}

We estimate $\epsilon_{H\alpha}\approx0.01$ for SN\,2018evt, which is comparable with the canonical value of $\epsilon_{\rm H_{\alpha}} = 0.05$ assumed in the literature \cite{Salamanca98,Taddia13}. 
Additionally, our computed $\frac{L_{\rm H\alpha}}{L_{\rm Opt}} \approx 0.07$ is also in general agreement with that reported in Figure 7 of \cite{Yang22}.

\subsection{4 The dust contributions of host galaxies by SNIa-CSM events}

The far-infrared (FIR) observations of the elliptical/lenticular (E/S0) galaxies by the Herschel Space Observatory suggest that the typical dust mass of such galaxies spans 10$^{4}$ to 10$^{7}$\,M$_{\odot}$ \cite{SmithMWL12,Finkelman10}. While average dust mass found in all types of galaxies in the local universe gives 10$^{5.21\pm0.09}$\,M$_{\odot}$.
Many dwarf elliptical galaxies exhibit dust masses less than 10$^{5}$\,M$_{\odot}$ \cite{DeLooze10}.

Based on our optical-to-MIR observing campaign on SN\,2018evt extended to day $\approx +1000$, we suggest that a total amount of 
$\approx$0.01\,M$_{\odot}$ newly-formed dust formed in the post-shocked region of the CDS. 
As the SN ejecta cools, more cold dust 
can be expected to form as illustrated in {\bf Figure 6}. We remark that the $0.01$ \Msun \ new dust formed in SN~2018evt is estimated 
only for warm dust based on the 3.6 and 4.5\,$\mu$m observations by NEOWISE and $Spitzer$. 
If we assume it estimates the typical mass of the warm dust formed in SNe Ia,  
the total mass of the newly-formed dust could be higher by a factor of $\sim$10 if the bulk of the dust cools below $\sim$30\,K.
The studies based on the observations of the Infrared Astronomical Satellite (IRAS) suggest $\sim$ 90\% of the dust in galaxies was missed by IRAS as IRAS or $Spitzer$ is sensitive to warm dust \cite{Devereux90}.

Depending on the detailed physical conditions, the timescale of the grain destruction could be as long as a few Gyr based on the revised self-consistent models on dust destruction efficiency  of SNe \cite{Priestley21}, and other cases \cite{DeLooze17}. 
Moreover, the rate of SN\,Ia per unit mass decreases as the stellar mass of the galaxy increases (e.g., Figure 5 in \cite{Graur2015:Ia:rate}). 
In particular, SN\,Ia rates of $6 \times 10^{-13}$ and $6 \times 10^{-14}$\,M$\odot^{-1}$\,year$^{-1}$ are estimated for galaxy stellar masses of 10$^{9}$ and 10$^{11}$\,M$\odot$, respectively. 
Assuming 0.1 \Msun \ of cold dust ($< 30$ K) is produced per SNIa, SNe Ia can produce on the order of $10^5$ - $10^6$ \Msun \ of dust for typical elliptical galaxies. Given the uncertainties in dust mass in ellipticals ($\rm M_d \le 10^5$ to $ 10^7$ \Msun), SNe Ia can be responsible for $10-100$\% of all the dust in elliptical galaxies. Considering that the SNIa-CSM rate is about 0.02 - 0.2\% of all SNe Ia \cite{Sharma23}, the dust from SNeIa-CSM may be proportionally lower than the above estimate for SNe Ia and can not be the dominant source of dust in elliptical galaxies. We note that the effect of galaxy merging is also a dust source in E/S0 galaxies as E/S0 galaxies can capture younger galaxies together with their dust. The captured dust is usually distributed in a thin disk, but dust is also present in a diffuse environment (e.g., \cite{Goudfrooij96}). SNIa-CSM contribution may also explain the diffuse dust.
%\clearpage
\section*{Figures and Extended Data}
\begin{figure*}[t]
\centering
    \includegraphics[width=12.5cm]{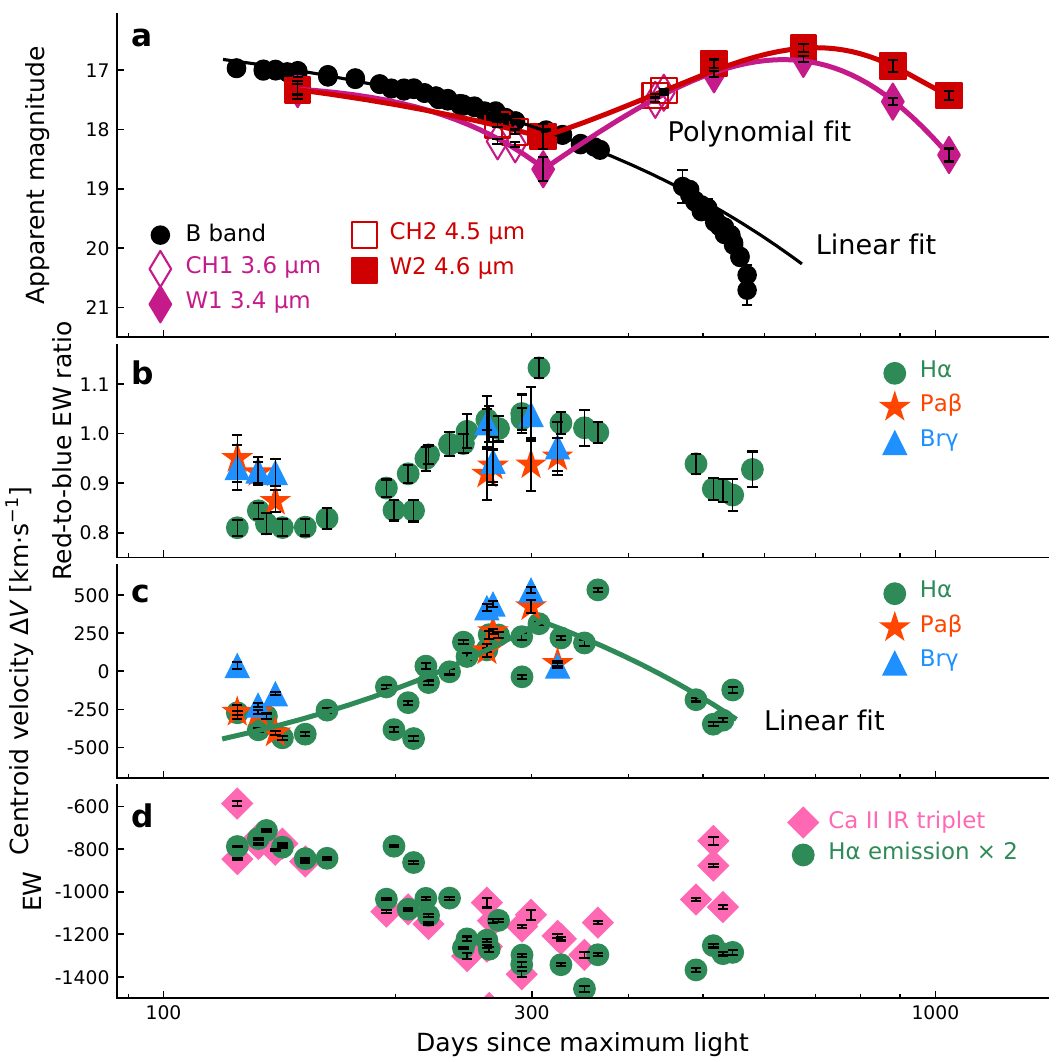}
    \caption{{\bf Evidence of the presence of dust in SN~2018evt.} 
    Panel (a) presents the MIR and the $B$ band light curves (black dots) of SN 2018evt. All phases are given relative to the estimated $B-$band maximum at MJD 58352. 
    The \textit{Spitzer} and NEOWISE observations are shown by purple diamonds and red squares as labeled. 
    The purple and red curves fit the MIR band 1 and band 2 photometry before and after day $+310$, separately.
    The black line fits the linearly-fading $B-$band photometry before day $\sim$400, with a decline rate of 0.624$\pm$0.006\,mag 100\,day$^{-1}$. 
 Panel (b) shows the red-to-blue EW ratios of H$\alpha$ (green circles), Pa$\beta$ (red stars), and Br$\gamma$ (blue triangles) lines.   
Panel (c) shows the evolution of the flux-weighted centroid velocities $\Delta V$ of H$\alpha$, Pa$\beta$, and Br$\gamma$ lines labeled with the same symbols as panel (b). The $\Delta V$ of H$\alpha$ measured before and after day +310 are fitted with separate linear functions as displayed by the two green line segments.
Panel (d) shows the evolution of the equivalent width (EW) of the Ca II IR triplet ({\bf Extended Data Table 2}) and H$\alpha$ lines. For the purpose of the presentation, the EW of H$\alpha$ has been multiplied by a factor of 2. 
The error bars shown
represent $1-\sigma$ uncertainties of magnitudes, EW ratio, centroid velocity, and EW.\\
    }
    \label{fig:whole}
\end{figure*}

\begin{figure*}[ht]
    \centering
    \includegraphics[width=15cm]{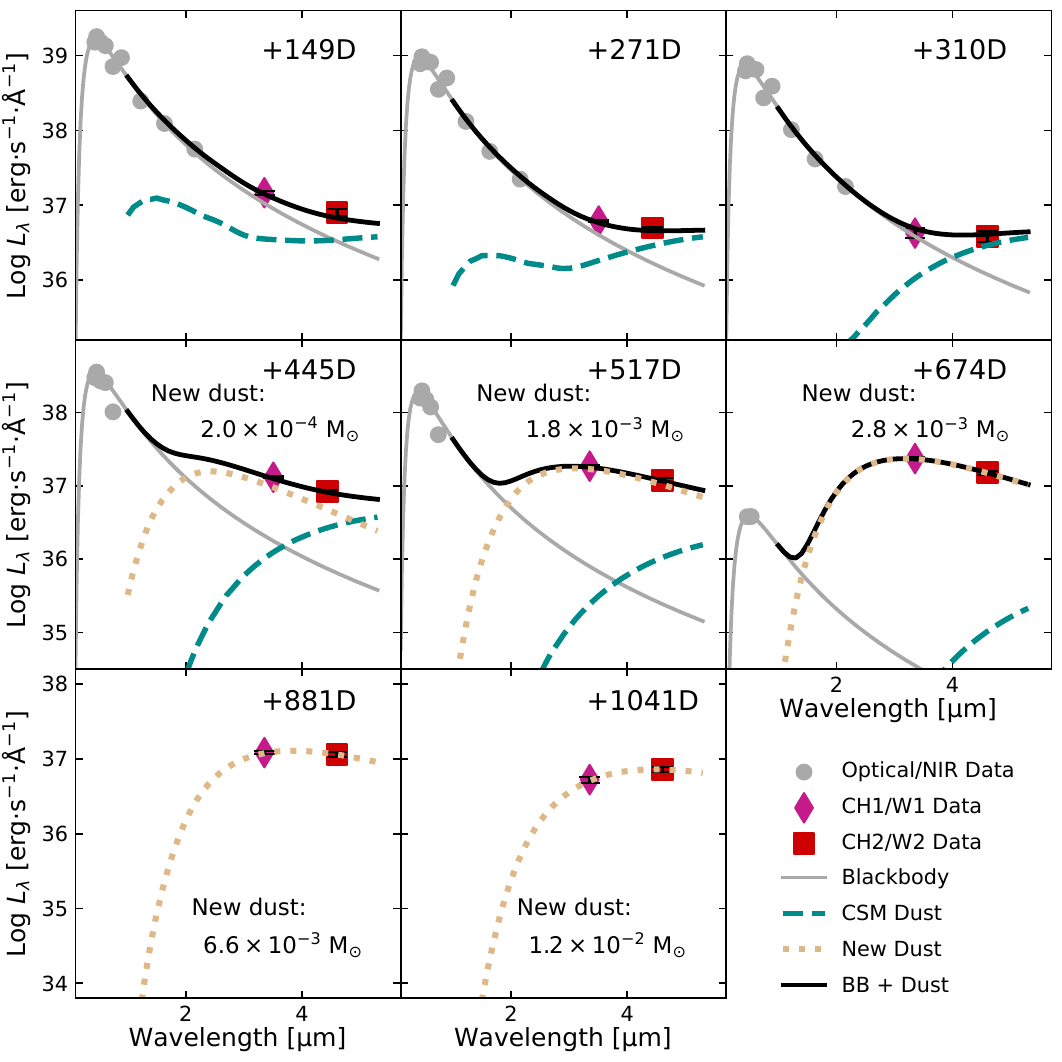}
    \caption{{\bf SED fitting of SN\,2018evt at the rest-frame wavelength.}
    The optical-to-NIR $BVgriJHK_{s}$ data are fitted by a single blackbody before day $+$674.
    Emissions from the CSM dust calculated from our double-shell model (see Methods 2 and {\bf Figure 4} for more details) are illustrated by cyan-dashed lines. Emissions from the newly-formed dust are shown as yellow-dotted lines. Note that the thermal emission of the newly-formed dust becomes progressively more dominant over time since day $+$445.
    The error bars shown
represent $1-\sigma$ uncertainties of the monochromatic luminosities.
}
    \label{fig:sed}
\end{figure*}
\clearpage

\begin{figure*}
    \centering
    \includegraphics[width=15cm]{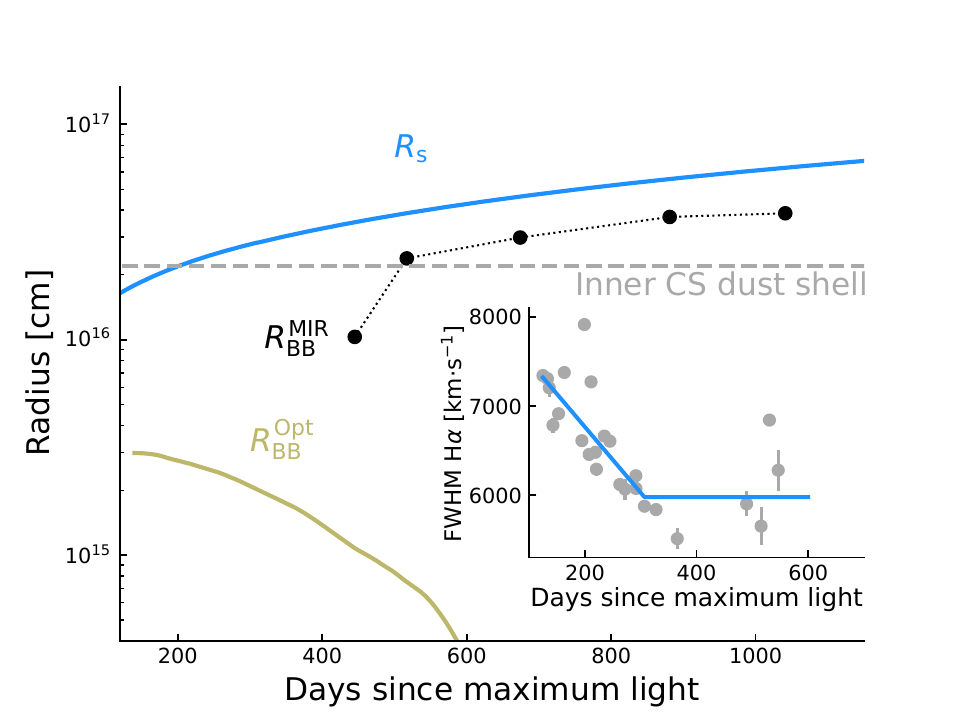}
    \caption{{\bf Time evolution of the different radii.} 
    The blackbody Photospheric radius $R_{\rm BB}^{\rm Opt}$， and the blackbody dust radius $R_{\rm BB}^{\rm MIR}$ are derived by fitting a blackbody spectrum to the optical-to-NIR luminosity and the MIR-flux excesses, respectively. The latter is displayed in {\bf Figure 4}, and the associated temperature of the newly-formed dust can be seen from the inset of {\bf Figure 6}. The horizontal gray-dashed line indicates the inner radius of the inner shell of the CSM ($2.2 \times 10^{16}$\,cm) in the double-shell model. The inset presents the temporal evolution of the FWHM width of the broad H$\alpha$ line. Two blue line segments present linear fits of the data before and after day $+$310, respectively.
    The shock radius $R_{s}$ was derived by equation (1) in Methods 2 by assuming that the shock velocity was 10000\,km\,s$^{-1}$ before the first observation at day $\approx +$120, and approximated by the FWHM width of the broad H$\alpha$ afterward.
    The error bars shown
represent $1-\sigma$ uncertainties of FWHM.
}
    \label{fig:radius}
\end{figure*}
\clearpage

\begin{figure*}[ht]
    \centering
    \includegraphics[width=15cm]{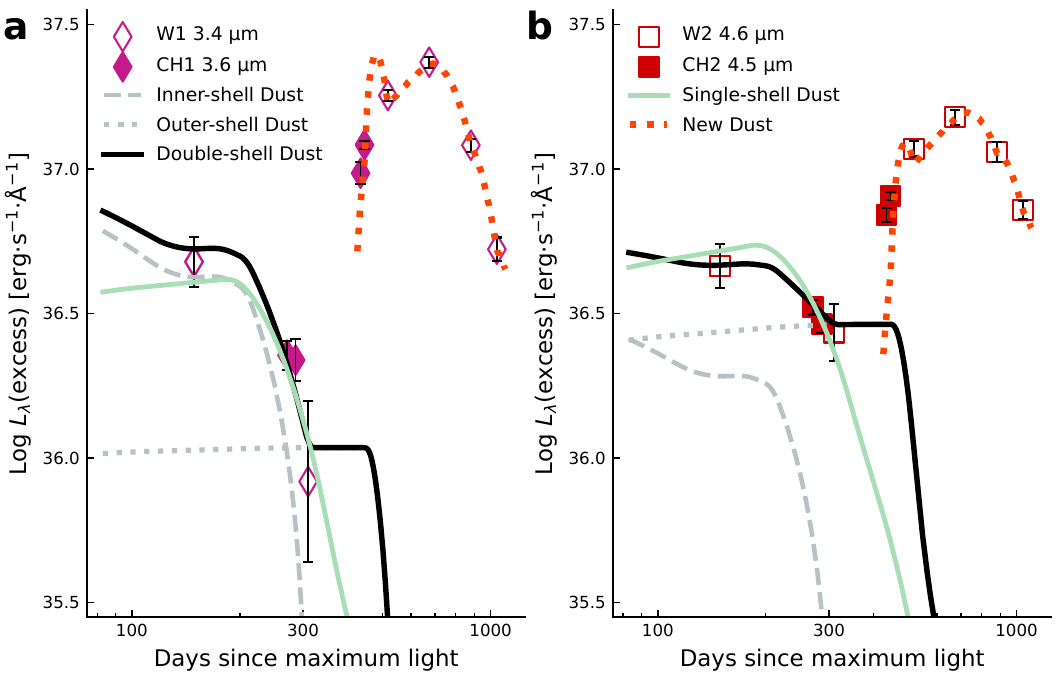}
    \caption{ {\bf Modeling to the MIR flux excesses of SN\,2018evt.} 
     The single-shell CSM dust model (green-solid line) with an inner radius of $2.6 \times 10^{17}$\,cm can fit well the declining flux excess in the MIR at day $\lesssim +310$ and infer a flatter power-law index of the dust density $s=1.15$. In the case of the steady wind mass loss $s=2.0$ for the double-shell model (black-solid line),
    dust grains within the inner shell at $2.2 \times 10^{16}$\, cm were continuously destroyed by the expanding forward shock between days $\sim$ $+$200 and $+$310, causing a monotonically-decreased flux excess in the MIR (gray-dashed line).
    The presence of an outer CSM dust shell with an inner radius of $6.0 \times 10^{17}$\,cm is necessary to account for the time evolution of the flux excess before day $+$310 (gray dotted line).
    The prominent rise of the MIR flux excess of SN\,2018evt after day $+$310, which cannot be explained by the thermal emission of any pre-existing dust content, demands a substantial amount of new dust to form promptly in the post-shock regions (red-dotted line). 
    The error bars shown
represent $1-\sigma$ uncertainties of monochromatic
luminosity excesses.
    }   
    \label{fig:model}
\end{figure*}
\clearpage

\begin{figure*}
    \centering
    \includegraphics[width=16cm]{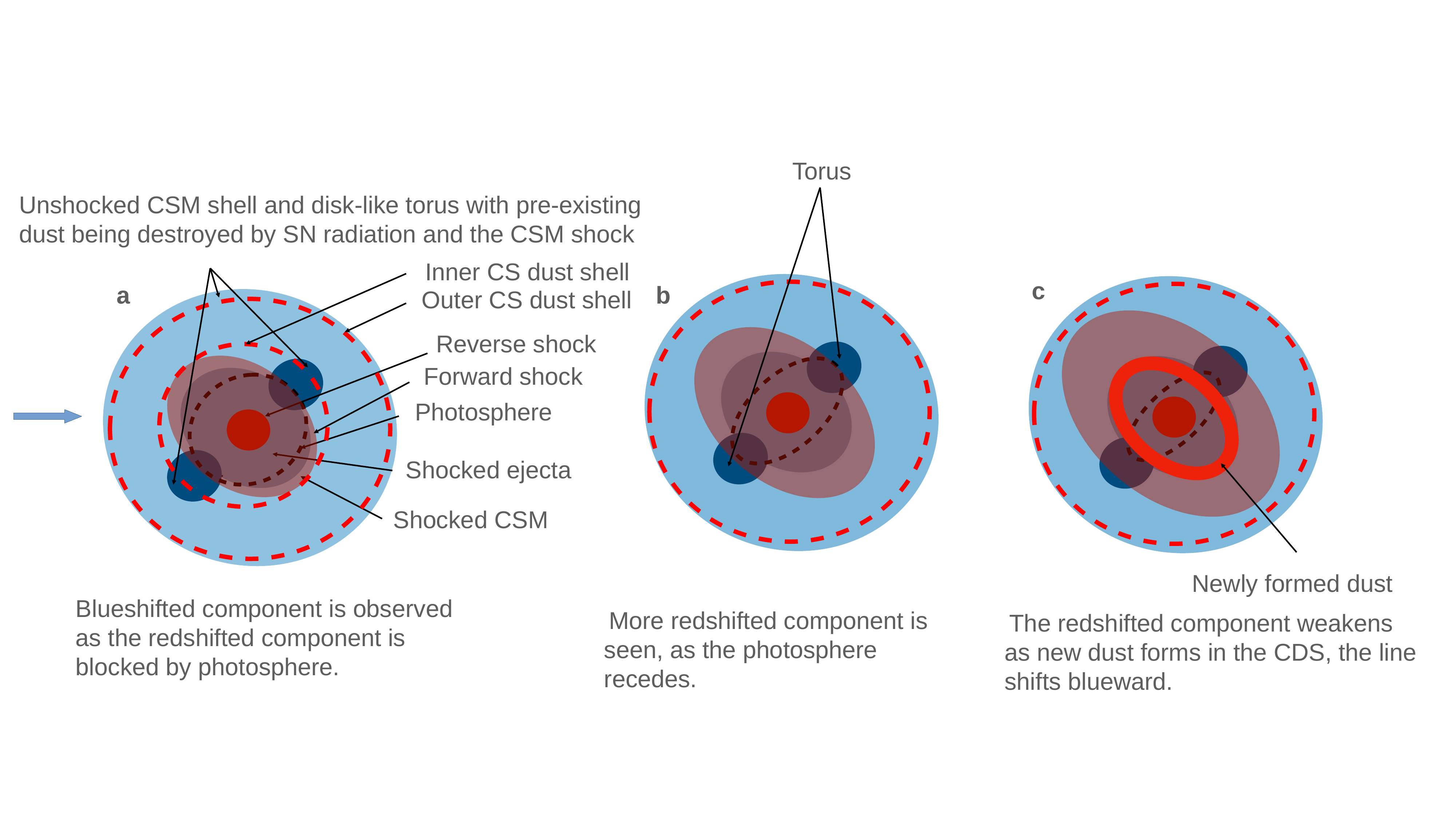}
    \caption{{\bf Schematic sketches of SN~2018evt at the different phases. }The blue arrow marks the viewing direction.
    The inner and outer CSM dust shells of our double-shell model are shown as red-dashed circles. The double-shell CSM model to describe the SED evolution of SN\,2018evt at day $\lesssim +310$ suggests inner radii of $2.2 \times 10^{16}$\,cm and $6.0 \times 10^{17}$\,cm for the inner and the outer shells, respectively. The single-shell CSM dust model infers an inner radius of $2.6 \times 10^{17}$\,cm. The sketches can represent the single-shell model after deleting the inner CS dust shell in panel (a).
    The brown-dashed ellipses approximate the location of the receding photosphere as the SN ejecta expands over time. 
    Panel (a) depicts the geometric configuration before day $+$310 when the redshifted component of the H$\alpha$ line (shown as solid blue ellipses) is blocked by the photosphere, producing  blueshifted line profiles. 
    As the SN ejecta expands and the photosphere recedes over time, more redshifted emission becomes revealed, resulting in a redward evolution of the line centroid as seen in {\bf Figure 1} c. When new dust forms at the post-shock CDS (illustrated by the thick-solid line), the redshifted side of H$\alpha$ is blocked again leading to blueward evolution of the line profile again. 
    }

    \label{fig:schematic}
\end{figure*}

\clearpage

\begin{figure*}
    \centering
     \includegraphics[width=16cm]{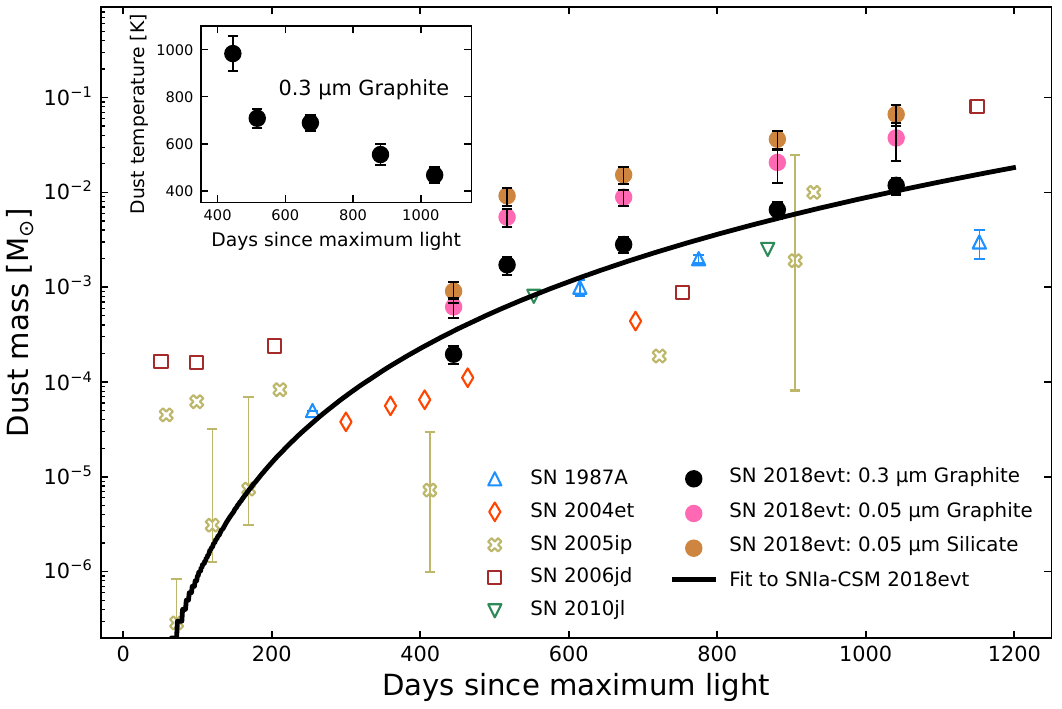}
    \caption{{\bf Temporal evolution of the mass of the newly-formed dust.} 
    As shown by the black line, the mass of the newly-formed dust of SN\,2018evt can be well fitted by a power law, i.e., $M_{\rm d} \propto t^{4}$ for 0.3\,$\mu$m graphite grains. 
    The dust masses calculated for graphite and silicate particles of radius 0.05\,$\mu$m are also presented.
     The inset traces the temperature evolution of newly-formed graphite dust particles of radius $0.3\ \mu$m. 
     The dust masses estimated for Type IIP SNe 2004et and 1987A, and IIn SNe 2005ip, 2006jd, and 2010jl are also shown.
    {\bf The error bars shown
represent $1-\sigma$ uncertainties of masses, and temperatures.}
     }
    \label{fig:dust}
\end{figure*}

\clearpage

\appendix
\renewcommand\thefigure{Extended~Data~\arabic{figure}}
\setcounter{figure}{0}

\begin{figure*}
    \centering
     \includegraphics[width=14cm]{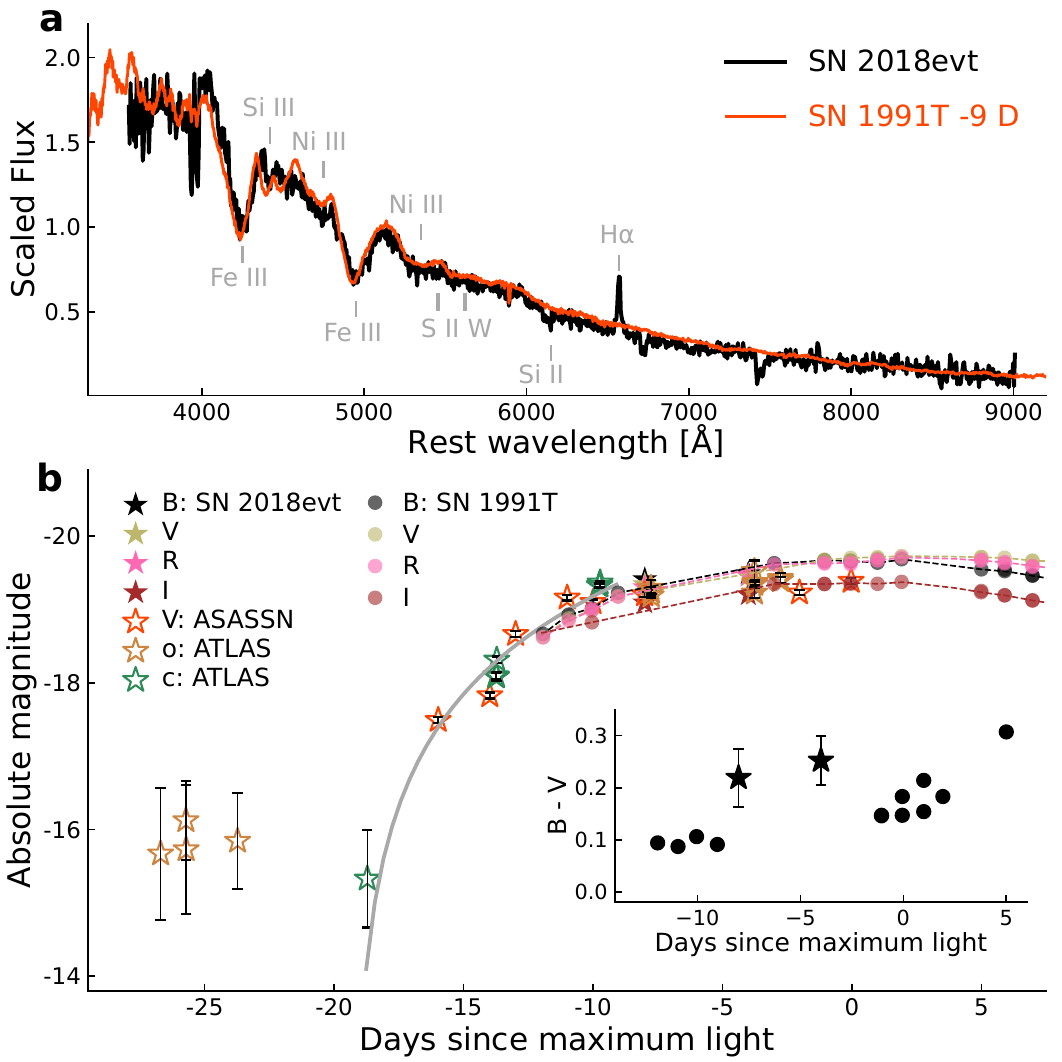}
    \caption{ {\bf The early-time comparisons of SNe\,2018evt, and 1991T.}
The spectrum of SN~2018evt (black curve \cite{class:18evt}) closely resembles the spectrum of SN\,1991T at day -9 (red curve) in panel (a) \cite{Filippenko92:91t, Mazzali95:91t}, which exhibits
the strong \ion{Fe}{III} $\lambda$4404, $\lambda$5129 absorptions and H$\alpha$ emission, visible \ion{Si}{III} $\lambda$4564 absorption, \ion{Ni}{III} blends around 4750 \AA, 5350 \AA, and weak \ion{S}{II} W and \ion{Si}{II} $\lambda$6355 as marked (e.g., \cite{Phillips22}).
Panel (b) compares the early-time photometry of SN~2018evt (stars), and SN~1991T (solid circles) \cite{Lira98}. 
A power law $f \propto (\tau + t_{r})^n$ is applied to fit the early $V-$ and $c-$band photometry, where $\tau =\frac{t-t_B^{max}}{s(1+z)}$ \cite{Ganeshalingam11_trise}, $t_B^{max}=58352$, $s=1.0$ for stretch value and $z=0.02523$ for the redshift of SN~2018evt.
The fitting yields a rise time $t_r=18.76\pm 0.24$ days, and a power-law index $n=1.57\pm0.07$ (gray curve). The estimated $t_r$ is consistent with the $V-$band rise time $t_r(V)=20.00\pm0.68$ days of SN~1991T/1999aa-like events \cite{Ganeshalingam11_trise}. The interpolations of $BVRI$ light curves of SN~1991T are shown in dashed curves. 
The inset in panel (b) compares the $B-V$ color curves between SNe~2018evt and 1991T, indicating a color difference $\lesssim0.1$ mag at similar phases.  
The corresponding Milkyway extinction is 0.05 mag \cite{SF2011} for SN~2018evt.  
{\bf The error bars shown
represent $1-\sigma$ uncertainties of magnitudes, and colors.}
}
\label{fig:cmp91t}
\end{figure*}  

\begin{figure*}
   \centering
    \includegraphics[width=15cm]{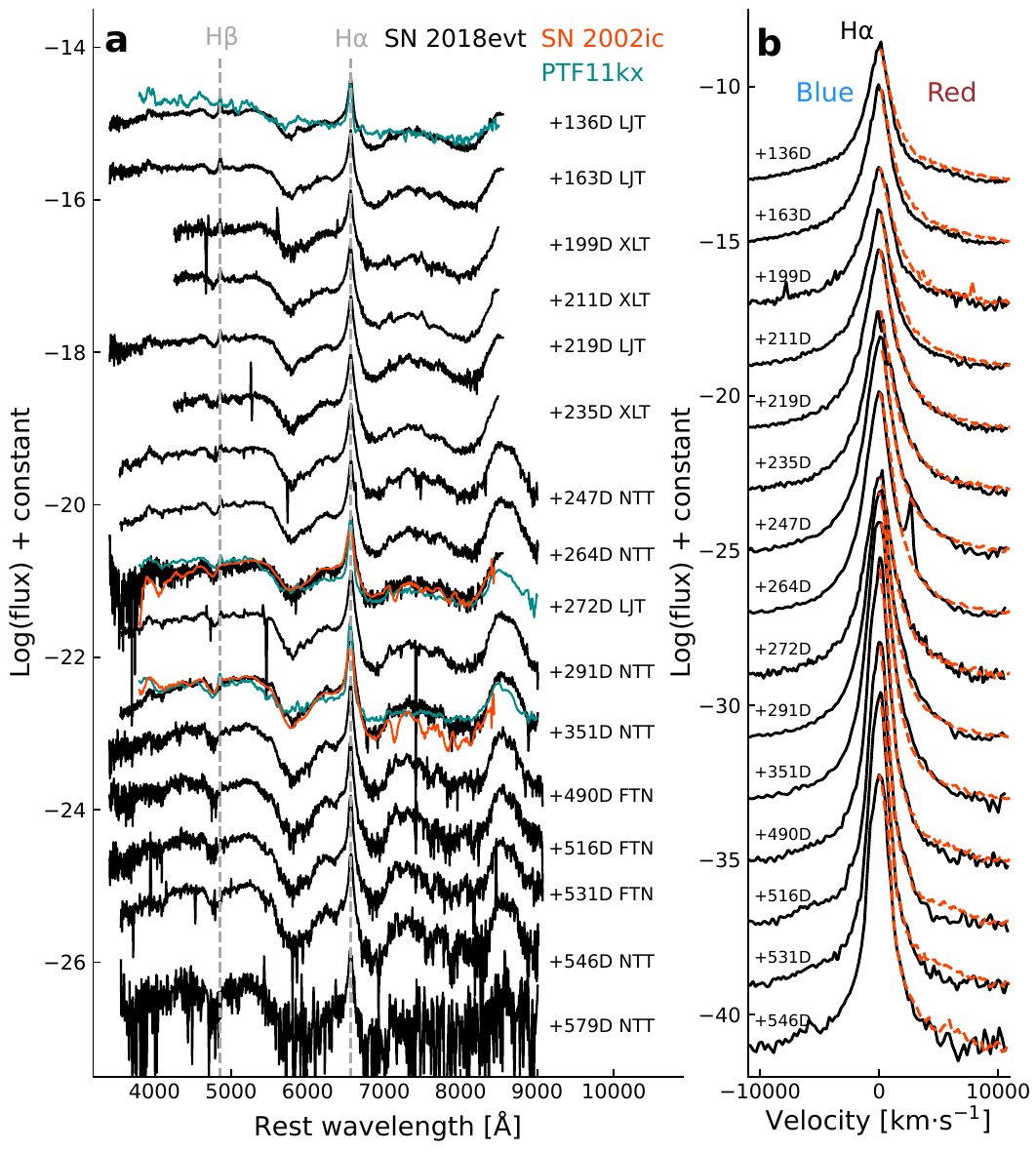}
    \caption{{\bf Optical spectra of SN 2018evt.} 
    Panel (a) shows optical spectra of SN 2018evt spanning from days $+136$ to $+579$ relative to $B$-band maximum. Phases and facilities are marked on the right. 
    Spectra of SN\,2018evt obtained at days $+$136, $+$272, and $+$351 are also compared to that of other Type Ia-CSM SNe (PTF11kx \cite{Dilday12,Silverman13:PTF11kx} and SN~2002ic \cite{Hamuy03:sn2002ic}) at similar phases.
    Panel (b) portrays the H$\alpha$ profile of SN~2018evt from panel (a). For each epoch, the red dashed line mirrors the spectral profile of the blue side across the peak flux of the intermediate H$\alpha$ (e.g., see {\bf Extended Data Figure 6} for two Gaussian fits to H$\alpha$). Its deviation from the red emission wing illustrates the time-variant asymmetry of the H$\alpha$ profile.
}
\label{fig:optsp}
\end{figure*}

%\clearpage

\begin{figure*}
    \centering
    \includegraphics[width=16cm]{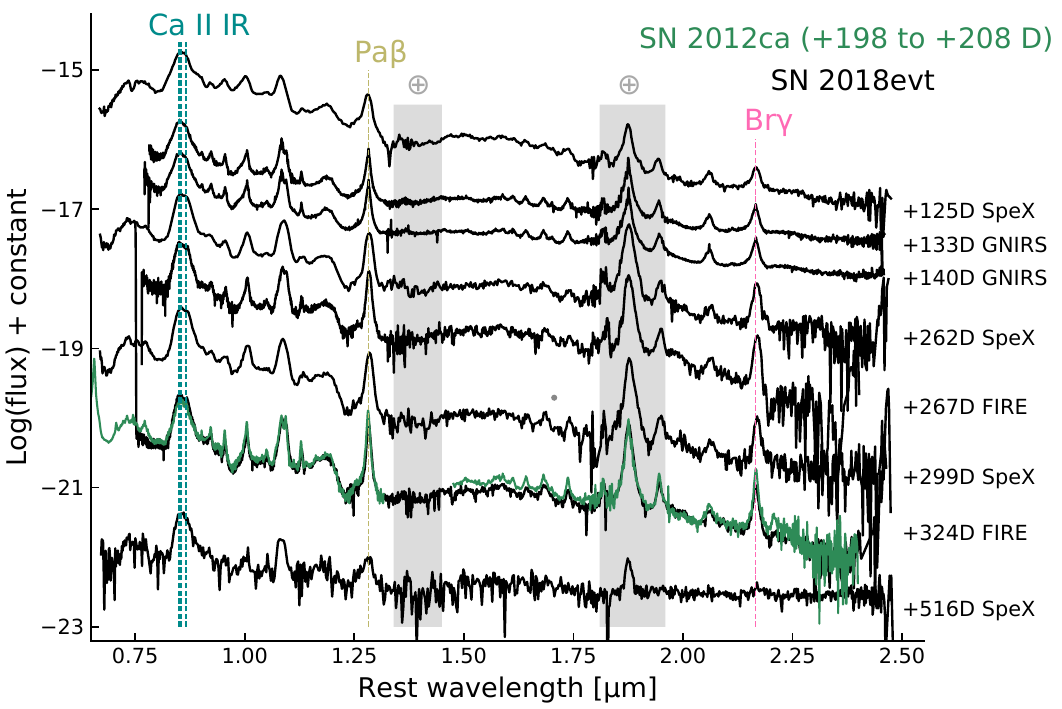}
    \caption{
    {\bf NIR spectra of SN 2018evt.} 
    Phases and facilities are marked on the right spanning from $\sim$ $+125$ to $+516$ days relative to the  $B$-band maximum. 
    Several most prominent lines are labeled. The near-IR spectrum of SN\,2012ca obtained at days $+198 - + 208$ is shown for comparison \cite{Inserra14}.
}
\label{fig:nirsp}
\end{figure*}

\begin{figure*}[ht]
    \centering
    \includegraphics[width=15cm]{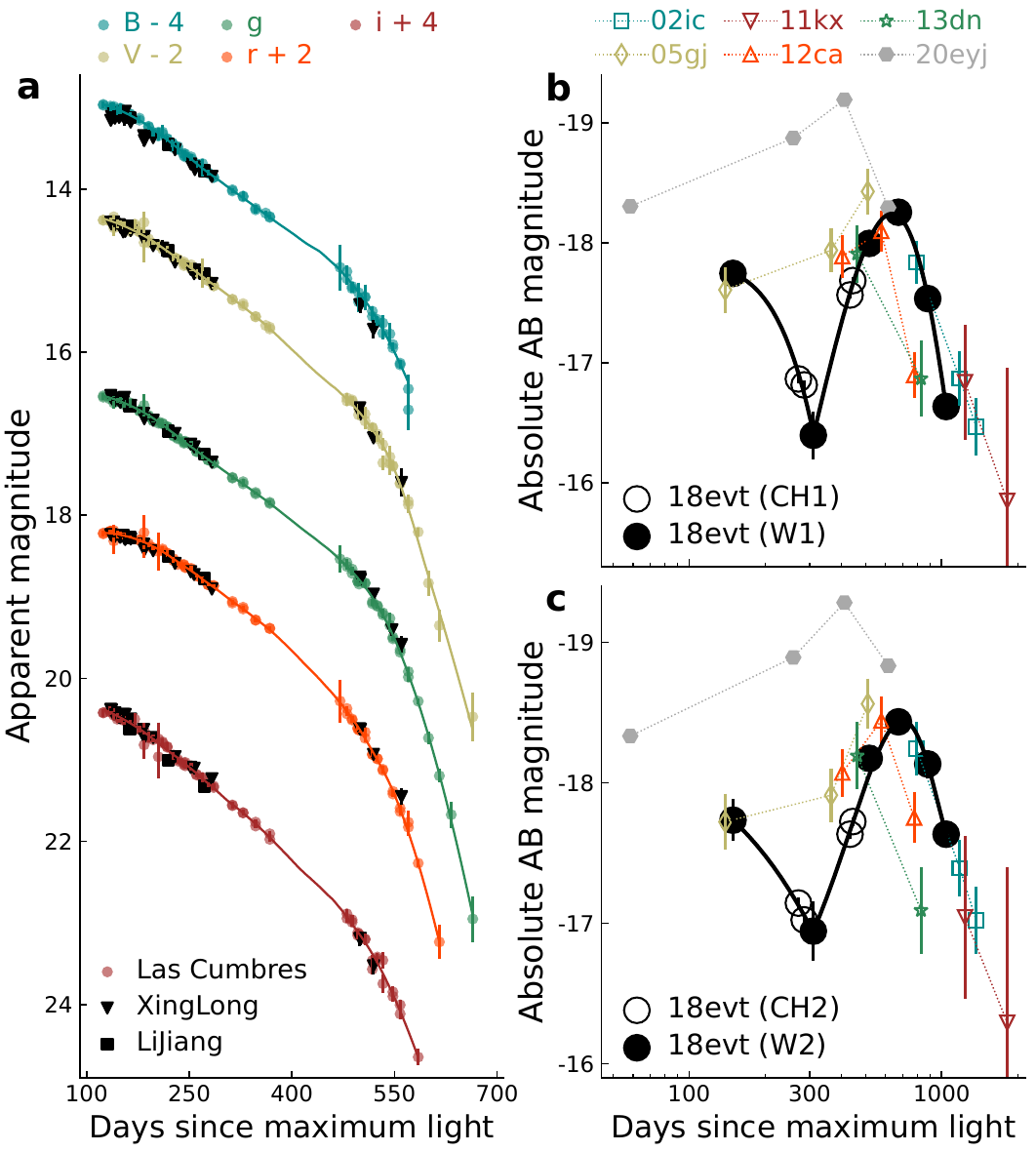}
    \caption{{\bf The optical and MIR light curves of SN\,2018evt.}
    Panel (a) The $BVgri$ band light curves of SN 2018evt. 
    Panels (b) and (c) present the $\sim$3.5\,$\mu$m (Spitzer $CH1$ and NEOWISE $W1$) and $\sim$4.6\,$\mu$m (Spitzer $CH2$ and NEOWISE $W2$) photometry of SN\,2018evt, respectively.
    Black-solid lines show polynomial fits to the light curves before and after day $+$310. 
    The MIR light curves of several other SNe\,Ia-CSM at similar phases are shown for comparison, including SNe~2002ic, 2005gj \cite{Fox13-05gjMIR}, PTF11kx \cite{Graham17}, 2012ca, 2013dn \cite{Szalai19,Szalai21}, and 2020eyj \cite{Kool23}.
    The error bars shown
represent $1-\sigma$ uncertainties of magnitudes.
}
    \label{fig:lc}
\end{figure*}

\begin{figure*}
    \centering
   \includegraphics[width=19cm]{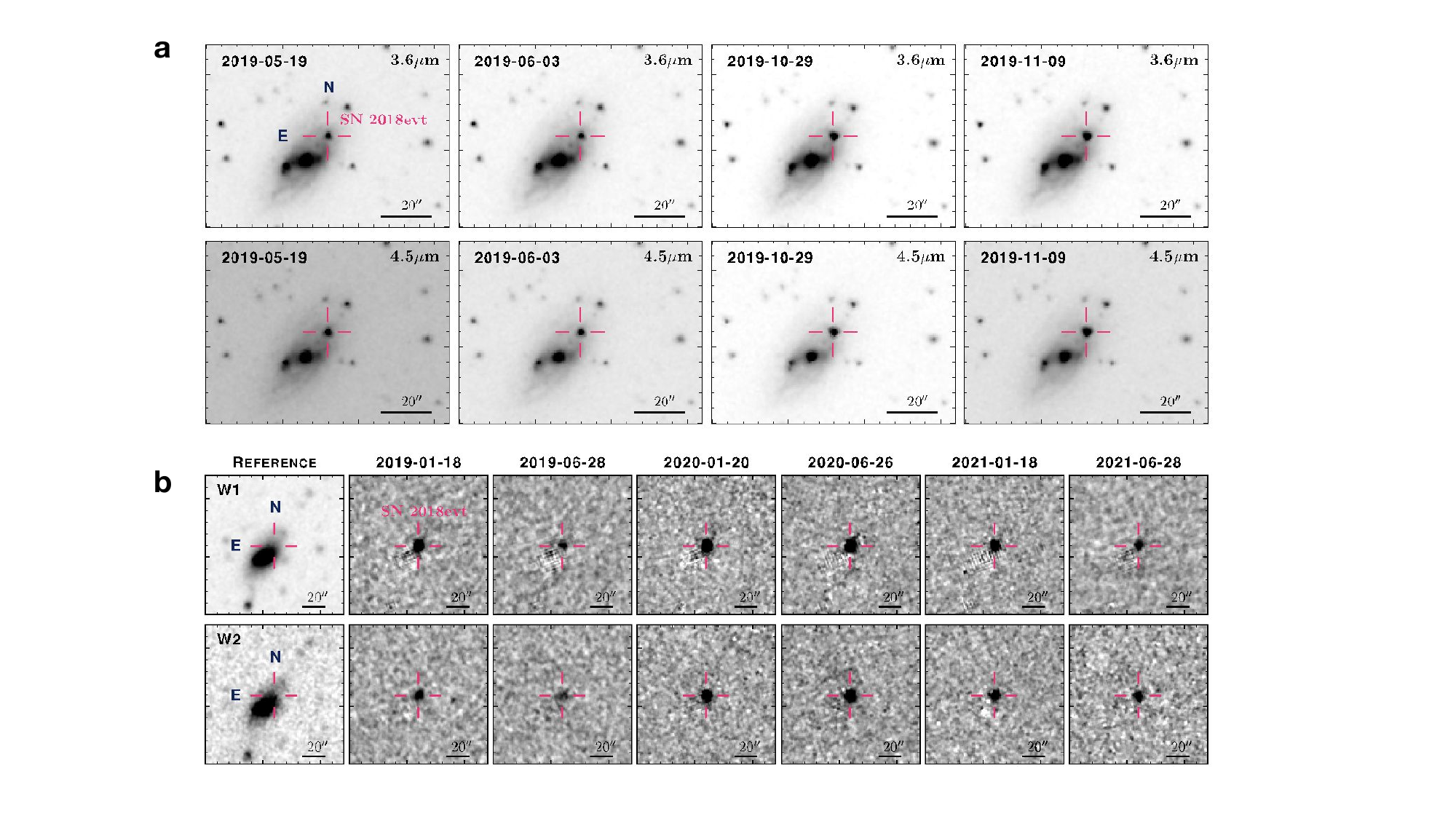}
   \caption{ {\bf The images of SN\,2018evt observed with Spitzer and NEOWISE.}
   The first and the second rows display the Spitzer $CH1$ (3.6\,$\mu$m) and $CH2$ (4.5\,$\mu$m)$-$band images obtained from 2019-05-19 (day $+$271) to 2019-11-09 (day $+$445), respectively.
   The third and the fourth rows present the NEOWISE $W1$ (3.4\,$\mu$m) and $W2$ (4.6\,$\mu$m) observations of the SN\,2018evt field, respectively. The left column shows the reference images constructed by co-adding the pre-SN exposures between January 2017 and January 2018. The reference-subtracted images obtained from 2019-01-18 (day $+$149) to 2021-06-28 (day $+$1041) are shown in the remaining subpanels as labeled. In each subpanel, the magenta cross indicates the location of SN\,2018evt. North is up, east is to the left.
}
    \label{fig:image}
\end{figure*}

\clearpage
\begin{figure*}
    \centering
    \includegraphics[width=16cm]{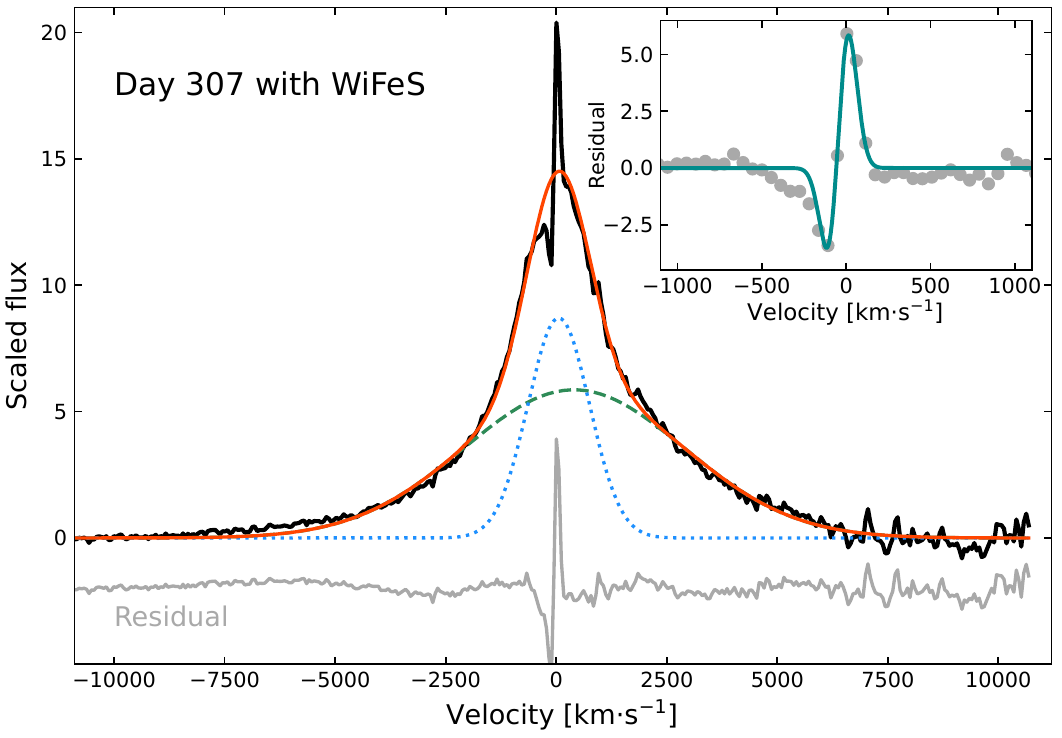}
    \caption{
{\bf The H$\alpha$ profile of SN~2018evt observed with WiFeS at day $+307$ fitted with two Gaussian functions.} 
The FWHM widths of the broad (cyan-dashed line) and intermediate (blue-dotted line) components yield 5877$\pm$32\,km\,s$^{-1}$ and 1643$\pm$12\,km\,s$^{-1}$, respectively. The red-solid curve gives the combination of these two components.
The bottom gray line represents the H$\alpha$ profile after subtracting the broad and intermediate components. An arbitrary offset has been applied to the residual spectrum for the purpose of presentation.
The inset provides a zoom-in view of the P-Cygni profile as displayed in the residual spectrum. A double-Gaussian component fit to the residual spectrum near the H$\alpha$ core is illustrated by the cyan curve. The location of the peak of the emission component suggests a redshift $z=0.02561\pm0.00019$, the location of the minimum of the absorption component measures a wind velocity $V_{w}=91\pm58$\,km\,s$^{-1}$.
}
\label{fig:wifes}
\end{figure*}

\clearpage

\begin{figure*}
    \centering
    \includegraphics[width=15cm]{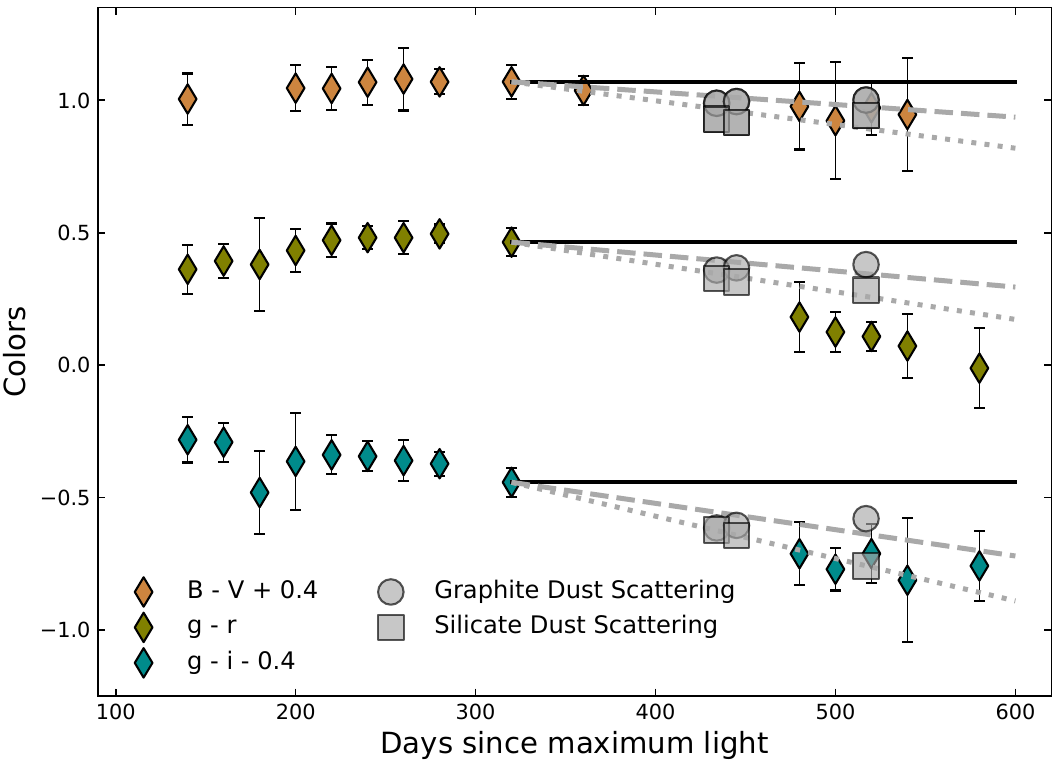}
    \caption{
   {\bf Galactic extinction-corrected $B-V$, $g-r$, and $g-i$ color curves of SN\,2018evt.} All colors were binned for 20 days to increase the signal-to-noise ratio.
   The colors predicted by the scattering of the newly-formed dust in the post-shock CDS are also presented by gray symbols as labeled. The calculation was carried out by assuming that the intrinsic colors of SN\,2018evt are identical to the values at day $\sim +310$, as indicated by the horizontal black line segments.
   The results of the 0.05\,$\mu$m graphite are shown by gray circles and linearly fitted by gray-dashed lines. The predicted colors of the 0.05\,$\mu$m silicate dust are presented by gray squares and linearly fitted by gray-dotted lines.
The error bars shown
represent $1-\sigma$ uncertainties of colors.
}
    \label{fig:color}
\end{figure*}

{\noindent \bf Extended Data Table 1 $\mid$  
 The MIR AB magnitudes of SN\,2018evt, and the parameters describe the newly-formed dust.} The $CH1$ and $CH2$ magnitudes were measured from the images taken by the Infrared Array Camera (without host subtraction) and the host-subtracted $W1$ and $W2$ magnitudes were measured from the observations by the NEOWISE reactivation mission. 
 The presented parameters are deduced for $a=0.3$\,$\mu$m, and $a=0.05$ \ $\mu$m graphite and silicate dust grains.
\begin{table}     
\centering
\rotatebox{90}{
\footnotesize
\scalebox{0.9}{
\begin{tabular}{ccccccrcrcr}
\hline
& & & & & \multicolumn{2}{c}{$0.3\ \mu$m Graphite} & \multicolumn{2}{c}{$0.05\ \mu$m Graphite} & \multicolumn{2}{c}{$0.05\ \mu$m Silicate} \\
UT Date &
MJD  & Phase&
$m_{AB,[CH1= 3.6\mu m]}$ & $m_{AB,[CH2= 4.5\mu m]}$   
   &Temperature 	
   &Dust mass
   &Temperature 	
   &Dust mass
   &Temperature 	
   &Dust mass \\
 (yy-mm-dd)    & & (day)  & (mag) & (mag) 
   & (K)
   & (M$_{\odot}$)
   & (K)
   & (M$_{\odot}$)
   & (K)
   & (M$_{\odot}$)\\
\hline
%date       MJD           ch1[3.6um]         ch2 [4.5um]
2019-05-19 & 58622.91 &$+270.91$ & $18.20\pm0.08$ & $17.92\pm0.08$ & & & & & & \\
2019-06-03 & 58637.50 &$+285.50$ & $18.25\pm0.08$ & $18.04\pm0.08$ & & & & & & \\
2019-10-29 & 58785.62 &$+433.62$ & $17.50\pm0.06$ & $17.43\pm0.06$ &$997\pm234$ &$1.4\pm0.7\times10^{-4}$ & $1034\pm259$ & $4.5\pm2.2\times10^{-4}$ & $1466\pm378$ & $7.4\pm3.0\times10^{-4}$\\
2019-11-09 & 58796.79 &$+444.79$ & $17.38\pm0.06$ & $17.34\pm0.06$ &$980\pm77$ &$2.0\pm0.4\times10^{-4}$  & $1026\pm89$ & $6.1\pm1.4\times10^{-4}$ & $1511\pm193$ & $9.1\pm2.2\times10^{-4}$ \\
\hline\\
UT Date  &
MJD  & Phase &
$m_{AB,[W1= 3.4\mu m]}$ & $m_{AB,[W2= 4.6\mu m]}$ & & & & & &\\
 (yy-mm-dd)   & & (day)  & (mag) & (mag) & & & & & &\\
\hline
2019-01-18 & 58501.21 &$+149.21$ & $17.32\pm0.09$ & $17.33\pm0.15$ & & & & & & \\
2019-06-28 & 58662.37 &$+310.37$ & $18.67\pm0.20$ & $18.12\pm0.21$ & & & & & & \\
2020-01-20 & 58868.60  & $+516.60$ & $17.07\pm0.05$ & $16.89\pm0.07$ &$706\pm39$ &$1.7\pm0.4\times10^{-3}$  & $727\pm43$ & $5.5\pm1.2\times10^{-3}$ & $918\pm68$ & $9.4\pm2.0\times10^{-3}$\\
2020-06-26 & 59026.36  & $+674.36$ & $16.81\pm0.05$ & $16.63\pm0.07$ &$689\pm32$ &$2.8\pm0.5\times10^{-3}$  & $707\pm34$ & $9.0\pm1.7\times10^{-3}$ & $892\pm58$ & $1.5\pm0.3\times10^{-2}$ \\
2021-01-18 & 59232.70  &$+880.70$  & $17.53\pm0.06$ & $16.93\pm0.10$ &$554\pm43$  &$6.6\pm1.3\times10^{-3}$ & $566\pm43$ & $2.1\pm0.8\times10^{-2}$ & $675\pm43$ & $3.6\pm0.8\times10^{-2}$ \\
2021-06-28 & 59393.48 & $+1,041.48$ &$18.43\pm 0.11$ & $17.43\pm0.08$ &$467\pm33$  &$1.2\pm0.2\times10^{-2}$ & $476\pm33$ & $3.7\pm1.6\times10^{-2}$ & $549\pm33$ & $6.7\pm1.6\times10^{-2}$\\
\hline
\end{tabular}
}
}
\end{table}

\clearpage

{\bf Extended Data Table 2: Log of the optical and NIR spectroscopic observations of SN\,2018evt.} The measured H$\alpha$ luminosity and the equivalent width of the \ion{Ca}{II} NIR triplet are also listed. \\
 $^a$ * marks the spectra that are already published in \cite{class:18evt,Yang22}.\\
 $^b$ Days since $B-$band maximum on MJD 58352 / 2018 August 22.\\
 $^c$ Uncertainty is derived and assumed that all spectra have 10\% flux uncertainty. Note that the distance is not included. Only the Milky Way extinction is corrected with $E(B-V)_{MW}=0.05$~mag, $R_V=3.1$. 

\begin{table}     
\centering
\footnotesize
\rotatebox{90}{
\begin{tabular}{lccccccccc}
\hline
{UT Date$^a$ } &
{MJD } & {Phase$^b$} &
{Resolution} & {Range} &
{Instrument/Telescope} &
{Exposure time} & {Airmass} &
log$L_{H\alpha}^c$ & $EW_{\rm Ca~II~IR~triplet}$\\
 (yy-mm-dd)          &          & (day)  &  (\AA)    & (\AA) &              & (s)  &  &
 (log(erg s$^{-1}$)) & (\AA)\\
           
\hline
2018-08-12* & 58343.00 & $-9.00$   & 15.8 & 3650-9200 & EFOSC2/NTT & 300  & 1.42& ... & ... \\
2018-12-24* & 58476.64 &   124.64 & $\sim$ 15.0 &3500-10000  & FLOYDS/2.0-m FTN & 1800 & 1.49    	 & $ 41.53 \pm0.01 $ 	&$-846.90\pm5.01$ \\
2019-01-01* & 58484.70 &   132.70 & $\sim$ 15.0 &3500-10000  & FLOYDS/2.0-m FTS & 1600 & 2.09    	 & $ 41.50 \pm0.01 $ 	&$-775.80\pm4.75$ \\
2019-01-04  & 58487.94 &  135.94 & 25.0 & 3400-9100 & YFOSC/LJT & 1350 & 1.38                   	 & $ 41.48 \pm0.01 $ 	& ... \\
2019-01-11* & 58494.58 &   142.58 & $\sim$ 15.0 &3500-10000  & FLOYDS/2.0-m FTN & 1600 & 1.56    	 & $ 41.52 \pm0.01 $ 	&$-774.40\pm4.64$ \\
2019-01-21* & 58504.64 &   152.64 & $\sim$ 15.0 &3500-10000  & FLOYDS/2.0-m FTN & 1600 & 1.18    	 & $ 41.52 \pm0.01 $ 	&$-857.92\pm5.11$ \\
2019-01-31  & 58514.89 &   162.89 & 25.0 & 3400-9100 & YFOSC/LJT & 1350 & 1.29                    	 & $ 41.52 \pm0.01 $ 	& ... \\
2019-03-04* & 58546.43 &   194.43 & $\sim$ 15.0 &3500-10000  & FLOYDS/2.0-m FTN & 1800 & 1.62    	 & $ 41.55 \pm0.01 $ 	&$-1093.59\pm6.55$ \\
2019-02-08  & 58550.75 &   198.75 & 15.0 & 4000-9000 & BFOSC/XLT & 3300 & 1.92                     	 & $ 41.45 \pm0.01 $ 	& ... \\
2019-03-17* & 58559.49 &   207.48 & $\sim$ 15.0 &3500-9250   & FLOYDS/2.0-m FTN & 1800 & 1.18    	 & $ 41.54 \pm0.01 $ 	&$-1082.88\pm6.53$ \\
2019-02-20  & 58562.75 &   210.75 & 15.0 & 4000-9000 & BFOSC/XLT & 3600 & 2.46                     	 & $ 41.46 \pm0.01 $ 	& ... \\
2019-03-28  & 58570.73 &   218.73 & 25.0 & 3400-9100 & YFOSC/LJT & 1500 & 1.31                    	 & $ 41.50 \pm0.01 $ 	& ... \\
2019-03-30* & 58572.49 &   220.49 & $\sim$ 15.0 &3500-10000  & FLOYDS/2.0-m FTN & 1800 & 1.17    	 & $ 41.51 \pm0.01 $ 	&$-1151.35\pm6.96$ \\
2019-04-13  & 58586.75 &   234.75 & 15.0 & 4000-9000 & BFOSC/XLT & 2700 & 1.63                     	 & $ 41.47 \pm0.01 $ 	& ... \\
2019-04-23* & 58596.66 &   244.66 & $\sim$ 15.0 &3500-10000  & FLOYDS/2.0-m FTS & 2700 & 1.31    	 & $ 41.49 \pm0.01 $ 	& ... \\
2019-04-26  & 58599.30 &   247.30 & 15.8 & 3650-9200 & EFOSC/NTT & 1500 & 1.51                    	 & $ 41.47 \pm0.02 $ 	&$-1302.33\pm14.06$ \\
2019-05-11* & 58614.39 &   262.39 & $\sim$ 15.0 &3500-10000  & FLOYDS/2.0-m FTN & 2699 & 1.19    	 & $ 41.44 \pm0.01 $ 	&$-1257.79\pm7.65$ \\
2019-05-13  & 58616.22 &   264.22 & 15.8 & 3650-9200 & EFOSC/NTT & 1499 & 1.28                    	 & $ 41.43 \pm0.02 $ 	&$-1545.90\pm16.17$ \\
2019-05-20  & 58623.69 &   271.69 & 25.0 & 3400-9100 & YFOSC/LJT & 2000 & 1.34                    	 & $ 41.40 \pm0.01 $ 	& ... \\
2019-06-09  & 58643.14 &   291.14 & 15.8 & 3650-9200 & EFOSC/NTT & 1800 & 1.24                    	 & $ 41.38 \pm0.02 $ 	&$-1387.15\pm14.73$ \\
2019-06-09* & 58643.37 &   291.37 & $\sim$ 15.0 &3500-10000  & FLOYDS/2.0-m FTN & 2700 & 1.40    	 & $ 41.39 \pm0.01 $ 	&$-1162.68\pm7.25$ \\
2019-06-24  & 58658.50 &   306.50 & ... & 3000-9500 & WiFes/ANU  & 1200 & ...                      	 & $ 41.47 \pm0.01 $ 	& ... \\
2019-07-15* & 58679.25 &   327.25 & $\sim$ 15.0 &3500-10000  & FLOYDS/2.0-m FTN & 3600 & 1.36    	 & $ 41.30 \pm0.01 $ 	&$-1221.34\pm7.56$ \\
2019-08-08  & 58703.04 &   351.04 & 15.8 & 3650-9200 & EFOSC/NTT & 2699 & 1.88                    	 & $ 41.26 \pm0.02 $ 	&$-1296.33\pm14.30$ \\
2019-08-22* & 58717.37 &   365.37 & $\sim$ 15.0 &3800-10000  & FLOYDS/2.0-m FTN & 3600 & 1.76    	 & $ 41.22 \pm0.01 $ 	&$-1145.15\pm7.78$ \\
2019-12-24  & 58841.62 &   489.62 & $\sim$ 15.0 &3500-10000  & FLOYDS/2.0-m FTN & 3600 & 1.54    	 & $ 40.76 \pm0.01 $ 	&$-1036.41\pm6.83$ \\
2020-01-19  & 58867.57 &   515.57 & $\sim$ 15.0 &3500-10000  & FLOYDS/2.0-m FTN & 3600 & 1.42    	 & $ 40.61 \pm0.01 $ 	&$-877.62\pm6.46$ \\
2020-02-03  & 58882.54 &   530.54 & $\sim$ 15.0 &3500-10000  & FLOYDS/2.0-m FTN & 3600 & 1.34    	 & $ 40.53 \pm0.01 $ 	&$-1071.95\pm7.56$ \\
2020-02-19  & 58898.29 & 546.29 & 15.8 & 3650-9200 & EFOSC/NTT & 2700 & 1.12                    	 & $ 40.37 \pm0.02 $ 	& ... \\
2020-03-23  & 58931.31 & 579.32 & 15.8 & 3650-9200 & EFOSC2/NTT & 2700 & 1.12                     	 & $ 40.16 \pm0.02 $ 	& ... \\
% NIR spectra
2018-12-24 &58476.65 & $+124.65$ &$\sim 18.3$ &7000-25000 & SpeX/IRTF & $150\times 10$ & 1.45        & ... & $-586.80 \pm   11.99$\\
2019-01-01 &58484.56 & $+132.56$ &$\sim 16.0$ &8000-25000 & GNIRS/Gemini North& $90\times20$ & 1.69  &... & $-752.82 \pm    5.06$\\
2019-01-08 &58491.63 & $+139.63$ &$\sim 16.0$ &8000-25000 & GNIRS/Gemini North& $90\times20$ & 1.28  &... & $ -803.16 \pm    5.53$\\ 
2019-05-11 &58614.40 & $+262.40$ &$\sim 18.3$ &7000-25000 & SpeX/IRTF & $150\times 10$ & 1.17        & ...&$-1051.25 \pm   22.74$\\
2019-05-16 &58619.22 & $+267.22$ &$\sim 24.0$ &7800-25000 & FIRE/Magellan Baade & 126.8$\times$12 & 1.26  &... &$-1136.63 \pm    8.47$\\
2019-06-17 &58651.30 & $+299.30$ &$\sim 18.3$ &7000-25000 & SpeX/IRTF & $150\times 10$ & 1.17        &... &$-1107.76 \pm   23.33$\\
2019-07-12 &58676.03 & $+324.03$ &$\sim 24.0$ &7800-25000 & FIRE/Magellan Baade & 126.8$\times$8 & 1.13 &... & $-1206.97 \pm    8.58$\\
2020-01-19 &58867.64 & $+515.64$ &$\sim 18.3$ &7000-25000 & SpeX/IRTF & $150\times 10$  & 1.18       &...&$-761.89 \pm   17.06$ \\

\hline
\end{tabular}
}
\end{table}

\clearpage

\clearpage
%reference
\section{Data availability} 
The data that support the findings of this study are openly available in Science Data Bank at \url{https://www.doi.org/10.57760/sciencedb.07968}\cite{Wang24:18evt:data} or \url{http://resolve.pid21.cn/31253.11.sciencedb.07968}. 
The global network photometry at Las Cumbres Observatory is also available in the figshare repository \url{https://doi.org/10.6084/m9.figshare.21543558}.
All spectra will also be made publicly available via Weizmann Interactive Supernova Data Repository. ATLAS \cite{Tonry18,Smith20} forced photometry service is available at \url{https://fallingstar-data.com/forcedphot/queue/}. All-Sky Automated Survey for Supernovae  (ASAS-SN \cite{Shappee14,Kochanek17}) sky patrol interface is available at \url{https://asas-sn.osu.edu/}. \textit{Spitzer} Heritage Archive (SHA) is available at \url{http://irsa.ipac.caltech.edu/applications/Spitzer/SHA/}. NEOWISE coadded images and ALLWISE source Catalog are available at \url{https://irsa.ipac.caltech.edu/applications/ICORE/}, and \url{https://irsa.ipac.caltech.edu/cgi-bin/Gator/nph-scan?submit=Select&projshort=WISE}. Pan-STARRS database \cite{Tonry12} is available at \url{https://catalogs.mast.stsci.edu/panstarrs}.

\section{Code Availability} 
SFFT \cite{Hu22:SFFT} package used for image subtraction are publicly available at \url{https://github.com/thomasvrussell/sfft}. IRAF soft used for spectra reduction is available at \url{http://iraf.net/}. BANZAI automatic pipeline used for image reductions at Las Cumbres Observatory is available at \url{https://lco.global/documentation/data/BANZAIpipeline/}. SExtractor \cite{sextractor} and PSFEx \cite{Bertin06} softwares used for photometry are available at \url{https://www.astromatic.net/software/sextractor/}, and \url{https://www.astromatic.net/software/psfex/}.

\section{Acknowledgements}
This work is sponsored (in part) by the Chinese Academy of Sciences (CAS), through a grant to the CAS South America Center for Astronomy (CASSACA) in Santiago, Chile. 
L.F.W. acknowledges support from National Science Foundation under grant No.-AST1817099. 
M.H. acknowledges support from the National Natural Science Foundation of China (grant no. 12321003), the Major Science and Technology Project of Qinghai Province (2019-ZJ-A10) and the Jiangsu Funding Program for Excellent Postdoctoral Talent. L.H. acknowledges support from Jiangsu Funding Program for Excellent Postdoctoral Talent and China Postdoctoral Science Foundation (Grant No. 2022M723372).
Y.Y. appreciates the generous financial support provided to the supernova group at U.C. Berkeley (PI: Alexei V. Filippenko) by Gary and Cynthia Bengier, Clark and Sharon Winslow, Sanford Robertson, and numerous other donors.
 T.W.C. acknowledges the Yushan Young Fellow Program by the Ministry of Education, Taiwan for the financial support.
X.W. acknowledges support from the National Natural Science Foundation of China under grants No. 11633002 and 11761141001. 
J.B., D.H., D.A.H., C.M., C.P., and E.P.G. from the GSP team at Las Cumbres Observatory are supported by NSF grants AST-1911225 and AST-1911151. 
P.H., E.Y.H., C.A., N.M., M.M.P., M.Sh., and M.D.S. from CSP-II group have been funded by the NSF under grants AST-1613426, AST-1613455, and AST-16133472.
J.C.W. is supported in part by NSF grant 1813825, by a DOE grant to the Wooten Center for Astrophysical Plasma Properties (WCAPP; PI Don Winget), and by grant G09-20065C from the Chandra Observatory. 
 L.G. acknowledges financial support from the Spanish Ministerio de Ciencia e Innovaci\'on (MCIN), the Agencia Estatal de Investigaci\'on (AEI) 10.13039/501100011033, and the European Social Fund (ESF) "Investing in your future" under the 2019 Ram\'on y Cajal program RYC2019-027683-I and the PID2020-115253GA-I00 HOSTFLOWS project, from Centro Superior de Investigaciones Cient\'ificas (CSIC) under the PIE project 20215AT016, and the program Unidad de Excelencia Mar\'ia de Maeztu CEX2020-001058-M.
Research by D.J.S. is supported by NSF grants AST-1821967, 1821987, 1813708, 1813466, 1908972, and by the Heising-Simons Foundation under grant \#2020-1864.
G.P. acknowledges support from the Millennium Science Initiative ICN12\_009.
M.Sh. is a visiting astronomer at the Infrared Telescope Facility, which is operated by the University of Hawaii under contract NNH14CK55B with the National Aeronautics and Space Administration.
T.E.M.B. acknowledges financial support from the Spanish 
Ministerio de Ciencia e Innovaci\'on (MCIN), the Agencia Estatal de 
Investigaci\'on (AEI) 10.13039/501100011033 under the 
PID2020-115253GA-I00 HOSTFLOWS project, from Centro Superior de
Investigaciones Cient\'ificas (CSIC) under the PIE project 20215AT016 and the I-LINK 2021 LINKA20409, and the program Unidad de Excelencia 
Mar\'ia de Maeztu CEX2020-001058-M.
M.N. is supported by the European Research Council (ERC) under the European Union’s Horizon 2020 research and innovation programme (grant agreement No.~948381) and by a Fellowship from the Alan Turing Institute.
M.G. is supported by the EU Horizon 2020 research and innovation programme under grant agreement No 101004719.
A.M.G. acknowledges financial support by the European Union under the 2014-2020 ERDF Operational Programme and by the Department of Economic Transformation, Industry, Knowledge,  and Universities of the Regional Government of Andalusia through the FEDER-UCA18-107404 grant. 
M.D.S. is funded in part by the Independent Research Fund Denmark (IRFD, grant number 10.46540/2032-00022B ).
S.Y. is supported by the National Natural Science Foundation of China under Grant No. 12303046.

This work makes use of observations from the Las Cumbres Observatory global telescope network.  
This publication makes use of data products from the Near-Earth Object Wide-field Infrared Survey Explorer (NEOWISE), which is a joint project of the Jet Propulsion Laboratory/California Institute of Technology and the University of Arizona. NEOWISE is funded by the National Aeronautics and Space Administration.
Based on observations obtained at the Gemini Observatory under programs GN-2018B-Q-136 (PI: Sand). Gemini is operated by the Association of Universities for Research in Astronomy, Inc., under a cooperative agreement with the NSF on behalf of the Gemini partnership: the NSF (United States), the National Research Council (Canada), CONICYT (Chile), Ministerio de Ciencia, Tecnolog\'{i}a e Innovaci\'{o}n Productiva (Argentina), and Minist\'{e}rio da Ci\^{e}ncia, Tecnologia e Inova\c{c}\~{a}o (Brazil). The data were processed using the Gemini IRAF package. We thank the queue service observers and technical support staff at Gemini Observatory for their assistance.
Based on observations collected at the European Organisation for Astronomical Research in the Southern Hemisphere, Chile, as part of ePESSTO and ePESSTO+ (the advanced Public ESO Spectroscopic Survey for Transient Objects Survey). ePESSTO observation was obtained under ESO program ID 199.D-0143 (PI: Smartt). ePESSTO+ observations were obtained under ESO programs ID 1103.D-0328 and 106.216C (PI: Inserra).
We have made use of the data with Magellan Baade/FIRE through CNTAC proposal IDs: CN2019B-8, CN2020B-23, (PI: Wang). 
We have also made use of the NASA/IPAC Extragalactic Database (NED) which is operated by the Jet Propulsion
Laboratory, California Institute of Technology, under contract with the National Aeronautics and Space Administration. 
\\
\\
\section*{Author contributions} 
L.Z.W. and L.F.W. wrote the manuscript. 
L.Z.W. planned the FIRE observations, reduced the NEOWISE and Spitzer images, compiled the whole dataset, and led the photometric, spectroscopic, and physical analysis. 
M.H. conducted the SED modeling of the data and contributed to the photometric data analysis. 
L.F.W. initiated the project and contributed to physical interpretation.
Y.Y. read the manuscript very carefully, offered very helpful comments to polish the paper, and also contributed to physical interpretation. 
J.Y. carried out the subtraction of the image through SFFT and reduced images from the global network. 
H.G. provided helpful comments on the dust in SNe. 
S.C. obtained the Spitzer images. 
L.H. helped with the subtraction of the NEOWISE image through SFFT.
G.P. and L.Y. helped with the FIRE observations.
C.C. helped with the calibration of the Spitzer images. 
D. B., P.H., and J.C.W. provided very helpful comments that improved the paper.
J.B., D.H., D.A.H., C.M., C.P., and E.P.G. from GSP group provided the extensive $BVgri$-band photometry and FLOYDS spectra. 
P.H., E.Y.H., C.A., N.M., M.M.P., M.Sh., and M.D.S. from CSP group provided FIRE and SpeX spectra. 
D.J.S.~provided the GNIRS spectra. 
T.W.C, L.G., J.P.A., M.G., C.I., A.M.G., T.E.M.B., M.N., J.P.G., M.S., S.S., S.Y., and D.R.Y. from ePESSTO group provided NTT spectra. 
J.M., X.W., H.L., H.S., and X.Z. provided TNTS photometry and spectra. 
J.Z. provided LJT photometry and spectra. 
S.A.U.~provided a WiFeS spectrum.

\section*{Competing interests} 
The authors declare no competing interests.

%\section*{Extended Data}

%\bibliography{sne}

\end{document}